	\documentclass[12pt]{iopart}
	\usepackage{iopams}  
	\expandafter\let\csname equation*\endcsname\relax
	\expandafter\let\csname endequation*\endcsname\relax
\usepackage{amsmath}
\usepackage{graphicx}
\usepackage[numbers, compress]{natbib}  
\usepackage{caption}
\DeclareFontFamily{U}{mathx}{\hyphenchar\font45}
\DeclareFontShape{U}{mathx}{m}{n}{
<-6> mathx5 <6-7> mathx6 <7-8> matha7
<8-9> mathx8 <9-10> mathx9
<10-12> mathx10 <12-> mathx12
}{}
\DeclareSymbolFont{mathx}{U}{mathx}{m}{n}
\DeclareMathSymbol{\bigplus}{\mathop}{mathx}{"90}
\DeclareMathSymbol{\bigtimes}{\mathop}{mathx}{"91}
\DeclareFontFamily{U}{FdSymbolA}{}
\DeclareFontShape{U}{FdSymbolA}{m}{n}{
    <-> s * [1] FdSymbolA-Book
}{}
\DeclareFontShape{U}{FdSymbolA}{m}{b}{
    <-> s * [1] FdSymbolA-Medium
}{}
\DeclareSymbolFont{fdsymbols}{U}{FdSymbolA}{m}{n}
\SetSymbolFont{fdsymbols}{bold}{U}{FdSymbolA}{m}{b}
\DeclareMathSymbol{\medtriangleright}{\mathbin}{fdsymbols}{86}
\DeclareMathSymbol{\medtriangleup}{\mathbin}{fdsymbols}{87}
\DeclareMathSymbol{\medtriangleleft}{\mathbin}{fdsymbols}{88}
\DeclareMathSymbol{\medtriangledown}{\mathbin}{fdsymbols}{89}
\DeclareMathSymbol{\medwhitestar}{\mathbin}{fdsymbols}{150}
\DeclareMathSymbol{\medsquare}{\mathbin}{fdsymbols}{127}
\DeclareMathSymbol{\medcircle}{\mathbin}{fdsymbols}{113}

\newcommand{\nRJ}{n^{\scriptscriptstyle \mathrm{RJ}}_\mathbf{k}}
\newcommand{\ERJ}{E^{\scriptscriptstyle \mathrm{RJ}}_\mathbf{k}}

\newcommand{\sgn}{\mathop{\mathrm{sgn}}}
\newcommand{\argmax}{\mathop{\mathrm{arg\,max}}}
\newcommand{\argmin}{\mathop{\mathrm{arg\,min}}}

\begin{document}

\title{Equilibria and condensates in Rossby and drift wave turbulence}
\date{}	
\author{Jonathan Skipp$^1$, and Sergey Nazarenko$^2$}

\address{$^1$ Centre for Complexity Science, University of Wawick, Coventry CV4~7AL, UK}
\address{$^2$ Universit\'{e}  C\^{o}te d'Azur,   CNRS, Institut de Physique de Nice, Parc Valrose, 06108 Nice, France}
\ead{j.m.skipp@warwick.ac.uk} 

\begin{abstract}
We study the thermodynamic equilibrium spectra of the Charney-Hasegawa-Mima (CHM) equation in its weakly nonlinear limit. In this limit, the equation has three adiabatic invariants, in contrast to the two invariants of the 2D Euler or Gross-Pitaevskii equations, which are examples for comparison. We explore how the third invariant considerably enriches the variety of equilibrium spectra that the CHM system can access. In particular we characterise the singular limits of these spectra in which condensates occur, i.e.\ a single Fourier mode (or pair of modes) accumulate(s) a macroscopic fraction of the total invariants. We show that these equilibrium condensates provide a simple explanation for the characteristic structures observed in CHM systems of finite size:  highly anisotropic zonal flows, large-scale isotropic vortices, and vortices at small scale. We show how these condensates are associated with combinations of negative thermodynamic potentials (viz.\ temperature). 
\end{abstract}




\section{Introduction}
\label{sec:intro}

It is well known that the turbulent flow of statistically isotropic ideal fluids in two dimensions (2D) conserves not only the total kinetic energy $E$  but also the total mean square vorticity, or enstrophy $\Omega$. The presence of these two invariants of motion places strong restrictions on how the flow may evolve, namely that any change must involve $E$ moving into large-scale motion of the flow while $\Omega$ concentrates in small-scale motions. Perhaps the most famous result in this direction is Kraichnan's dual cascade of 2D turbulence~\citep{kraichnan1967inertial}. However one of the earliest formulations is the simple and robust argument of Fj{\o}rtoft~\citep{fjortoft1953changes}, whose conclusion can be expressed as a principle involving the sign-definite global invariants of the flow: each such invariant is pushed by all the others towards the sector of Fourier space where its spectral weight is greatest~\citep{nazarenko2009triple, nazarenko2011waveturbbook}.

This statement, regarding the dynamics of 2D flow, is also reflected in the equilibrium statistical mechanics of 2D turbulence. Onsager~\citep{Onsager1949} studied point vortices in a spatially finite domain, and found that flow states with high energy involve like-signed vortices clustering together to form large-scale supervortices, and that these configurations are associated with a negative thermodynamic temperature. Later, Kraichnan~\citep{kraichnan1967inertial, kraichnan1975statistical} considered continuous vorticity fields by examining the 2D Euler equation with Fourier truncations at the small and large scale. He established that their statistical equilibria also predicted the large-scale (respectively small-scale) accumulation of energy (enstrophy), with the equilibrium being parameterised by a negative temperature (enstrophy temperature), see also review~\citep{kraichmonty1980}. Kraichnan also noted the analogy between the accumulation of energy into the largest scale in 2D turbulence, and the condensation of particles in a 2D Bose gas~\citep{kraichnan1967inertial}, directly terming the former as condensation of energy into the ``gravest mode''~\citep{kraichnan1975statistical}.

In planetary atmospheres and oceans, and in magnetically confined fusion plasmas, the slow mesoscale dynamics comprises of a quasi-2D flow constrained by the gradient of the Coriolis force in the geophysical case, and the diamagnetic drift in the plasma case. This gradient constraint, called the $\beta$ effect in the geophysical literature, breaks the isotropy of the system. Consequently in these systems it is observed that energy condenses not into the gravest isotropic mode, but rather into strongly anisotropic zonal flows: bands of alternating shear flows aligned latitudinally (geophysical), or in the poloidal direction of the confinement device (plasma),  and block the transport of energy and fluid across them~\citep{rhines1975waves, HasegawaMaclennanKomada1979, balk1990nonlocal, huang2001anisotropic, diamond2005zonal, shats2005spectral, galperin2019zonal}.

An important feature of such quasi-2D flows with a $\beta$ effect is that in addition to $E$ and $\Omega$, for systems with small nonlinearity there exists a third positive-definite adiabatic invariant of motion~\citep{balknazazakh1990structure, balknazazakh1991new, balk1991new}. This invariant explains the tendency for energy to condense into zonal jets by the following adaptation of the Fj{\o}rtoft argument. With three invariants, Fourier space is divided into three sectors, each of which is dominated by the cascade of its respective invariant. Such a division of 2D Fourier space into three sectors cannot be done isotropically, as it can in 2D turbulence when there are only two invariants. The specific form of the sectors is shown in figure~\ref{fig:InvDensityIsolines} below, with the gravest isotropic mode being occupied by the cascade of the third invariant. This ``throws'' the energy cascade onto zonal modes, with wavenumber components $k_x \ll k_y$; hence the third invariant has been termed zonostrophy~\citep{nazarenko2009triple}.

Such considerations are relevant in a non-equilibrium situation, either when the system is transiently evolving from an initial condition, or when there is persistent forcing and dissipation. In the present paper we offer an alternative explanation of the tendency for zonal modes to condense energy, based on the statistical equilibria of quasi-2D flows with a $\beta$ effect in the weakly nonlinear limit, and show that these too predict the condensation of energy into zonal flows. We study the Charney-Hasegawa-Mima (CHM) equation~\citep{charney1948scale, hasegawa1977stationary}, which is a simple model that describes free-energy-driven turbulence and zonal flow formation in both quasi-geostrophic geophysical flows, and drift dynamics in magnetically confined plasmas~\citep{HasegawaMaclennanKomada1979}. We examine the CHM equation in the weakly nonlinear limit, using the theory of wave turbulence (WT)~\citep{zakharov1992kolmogorovbook, nazarenko2011waveturbbook} to find the equilibrium spectra of flows where enstrophy, energy, and zonostrophy are all invariants of the motion.

Our approach is similar in spirit to Einstein's original analysis of Bose-Einstein condensation~\citep{einstein1925aQuantentheorie}, Kraichnan's analysis of the 2D Euler equation~\citep{kraichnan1967inertial, kraichnan1975statistical, kraichmonty1980}, and to the description by Connaughton et.\ al.\ \citep{connaughton2005condensation} of the condensation of classical waves obeying the Gross-Pitaevskii equation (GPE, also known as the nonlinear Schr{\"o}dinger equation), using WT theory. We note that the foregoing references derive equilibria using different ensembles. As is common in WT theory, including in~\citep{connaughton2005condensation}, we work here in the microcanonical ensemble, considering an isolated system evolving from an initial condition and finding a unique equilibrium through ergodic dynamics. This is in contrast to Kraichnan's analysis of 2D turbulence which took place in the grand canonical ensemble, although see~\citep{vanKan2021geometric} for a recent microcanonical treatment of the Fourier-truncated 2D Euler equation. 

It is worthwhile at this point to summarise the findings of~\citep{connaughton2005condensation} on the GPE in order to preview the WT approach we take here, and to specify concretely what the term ``condensation'' will refer to in this work. In the appendix we recapitulate results on infinite-sized GPE systems in 3D and 2D in slightly more detail, as well as apply the results of this work to the 2D GPE in finite-sized systems.

\subsection{Condensation outside and within the equilibrium spectrum}
\label{subsec:GPE_condensation}

The  GPE in the WT limit has two invariants: the total waveaction, or number of particles $\mathcal{N}$, and the leading order energy $\mathcal{E}$. Once initialised with this $\mathcal{N}$ and $\mathcal{E}$, the system will come to equilibrium on the so-called Rayleigh-Jeans (RJ) spectrum $\nRJ$ (see table~\ref{tab:CHM_GPE} for the functional form of $\nRJ$ and the invariants). Crucially, in three dimensions (3D) the integrals that define $\mathcal{N}$ and $\mathcal{E}$ on the RJ spectrum converge as the system size tends to infinity, meaning that $\nRJ$ can only accommodate a finite amount of particles and energy. There exists some nonzero ratio of energy per particle, i.e.\ temperature, below which the excess particles must collapse into a singular distribution at wavenumber $\mathbf{k}=0$; this distribution can absorb an arbitrarily large of the total number of particles in the system. This is the WT counterpart of Einstein's argument for condensation of a Bose gas~\citep{einstein1925aQuantentheorie}. Strictly speaking it is this singular distribution that is known as the true Bose-Einstein condensate~\citep{pitaevskiistringari2003book}.

By contrast in 2D, $\mathcal{N}$ diverges logarithmically at low wavenumber and so the RJ spectrum can absorb an arbitrary number of particles in the infinite box limit. Consequently no nonzero critical condensation temperature exists for a spatially infinite system, a manifestation of the Mermin-Wagner-Hohenberg theorem~\citep{MerminWagner, Hohenberg}. However for spatially finite 2D systems condensation is restored, in the sense of the largest-scale (``gravest'') mode containing a macroscopic fraction of the total $\mathcal{N}$~\citep{bagnato1991bose, safonov1998observation, during2009breakdown, HadzibabicDalibard2011_2Dbose, baudin2020classical}. Numerical demonstrations of finite-size condensation can also clearly be seen in~\citep{nazarenko2006wave, nazarenko2007freely}. It is in this sense that we refer to ``condensation'' in this paper: a macroscopic fraction of the invariants occupying a single fundamental mode \emph{within the equilibrium RJ spectrum}. As we will see, the meaning of the fundamental mode changes in the 2D anisotropic system we examine here. 

Finally, we note that in studying the GPE, we can take the ratio $\mathcal{E}/\mathcal{N}$ as the control parameter, as this ratio is accessible by specifying the initial spectrum. One can then take the chemical potential $\mu$ as the sole thermodynamic potential that characterises the equilibrium. The phase diagram for the GPE is thus one dimensional. In the CHM system we examine, the presence of an additional invariant requires a third thermodynamic potential to describe the equilibrium, and two ratios of invariants are needed as control variables. Thus the phase diagram for weakly nonlinear CHM turbulence requires two dimensions to visualise, and is much more intricate than the generic one dimensional phase diagram. We will see that it contains seven of the eight possible sign combinations for the three thermodynamic potentials, and that condensates are associate with the margins of the phase diagram, where at least one potential is negative.

\subsection{Organisation of this paper}
\label{organisation}

We start in section~\ref{sec:CHM} by recapitulating the WT of the CHM equation, defining its absolute and adiabatic invariants, and stating its RJ spectrum. We also make clear the correspondence betwen the CHM, the 2D GPE, and 2D Euler flow. Then in section~\ref{sec:RJdiverg} we consider the range of values that the thermodynamic potentials can take for a physical RJ spectrum, and enumerate the possible Fourier modes over which the RJ spectrum is singular, i.e.\ the fundamental modes where we expect condensates to form. In section~\ref{sec:condensation} we demonstrate that condensates, in the sense we have described above, do indeed form at the specified fundamental modes, and that these condensates do not survive in the infinite box limit, as expected from the Mermin-Wagner-Hohenburg theorem. In section~\ref{sec:CHMphasediag_discr} we plot the phase diagram for weak CHM turbulence.
We conclude in section~\ref{sec:conclusion}.

\section{Charney-Hasegawa-Mima equation}
\label{sec:CHM}

The CHM equation,
\begin{equation}
	\label{eq:CHM}
	\frac{\partial}{\partial t}\left(  \nabla^2 \psi  -  \rho^{-2}\psi \right)
		\,+\,
	\beta \frac{\partial\psi}{\partial x}
		\,+\,
	\left[ 
		\frac{\partial\psi}{\partial x} \frac{\partial \nabla^2 \psi}{\partial y}
		-\frac{\partial\psi}{\partial y} \frac{\partial \nabla^2 \psi}{\partial x} 
	\right]
	=0
	,
\end{equation}
describes the evolution of the stream function $\psi(\mathbf{x},t)$ in 2 spatial dimensions (in the plasma context $\psi$ also represents the electrostatic potential). 
$\rho$ is Rossby deformation radius (geophysics) or the ion gyroradius at the electron temperature (plasma).

In~\eqref{eq:CHM} we have adopted geophysical notation, in which $x$ is the zonal direction (west-east) and $y$ is the latitudinal (south-north) direction. In this convention when \eqref{eq:CHM} is interpreted in the plasma context $y$ is the direction of decreasing density and $x$ is the direction binormal to the magnetic field and the density gradient. $\beta$ measures the gradient of Coriolis parameter or plasma density gradient, which provides a restoring force to perturbations, allowing linear wave solutions of~\eqref{eq:CHM} in the absence of nonlinearity (terms in square brackets in~\eqref{eq:CHM} vanish). These are known as Rossby waves in geophysics and drift waves in the plasma context, and have dispersion relation
\begin{equation*}
	\label{eq:CHM_dispersion} 
	\omega_\mathbf{k} = -\beta  \rho^2 \frac{k_x}{1+\rho^2 k^2},
\end{equation*} 
where $\mathbf{k}=(k_x,k_y)$ is the wavenumber and $k=|\mathbf{k}|$. The WT we discuss in this paper is the turbulence of Rossby/drift waves, which comes about when the system is weakly nonlinear.

\subsection{Wave turbulence and the CHM kinetic equation}
\label{subsec:CHM_WT_KE}

We now describe the CHM equation in its WT limit.
We consider the system inside $\mathbb{T}^2_L$, a 2D periodic box of side length $L$, and decompose into Fourier modes 
$\psi_\mathbf{k}(t) = \int_{\mathbb{T}^2_L}\! \psi(\mathbf{x},t) \exp(-\mathrm{i}\mathbf{k}\!\cdot\!\mathbf{x}) \,\mathrm{d}\mathbf{x}$. We then assume the modes are statistically independent in phase and  amplitude, that the phases are uniformly distributed in $[0,2\pi)$, and we formally take the limits of an infinite box ($L \to \infty$), and weak nonlinearity. One can then derive a three-wave kinetic equation for the evolution of the waveaction spectrum~\citep{Longuet-Higgins1967resonant}
\begin{equation}
	\label{eq:CHM_nk}
	n_\mathbf{k} = \left(\frac{L}{2\pi}\right)^2 \frac{(1+\rho^2k^2)^2}{\beta\rho^4 k_x} \langle |\psi_\mathbf{k}|^2\rangle
\end{equation}  
where the angle brackets denote averaging over phases and amplitudes of the initial wave modes.
The kinetic equation is
	\begin{equation}\label{eq:CHM_kinetic}
	\frac{\partial n_\mathbf{k}}{\partial t} 
		=
	\int\limits_{k_{1x}, k_{2x} > 0} \! ( \mathcal{R}_{12\mathbf{k}} - \mathcal{R}_{\mathbf{k}12} - \mathcal{R}_{2\mathbf{k}1} )
		\,\mathrm{d}\mathbf{k}_1 \mathrm{d}\mathbf{k}_2,
\end{equation}
where
\begin{equation*}
	\mathcal{R}_{12\mathbf{k}} 
	= 
    2\pi  |V^\mathbf{k}_{12}|^2  \delta^\mathbf{k}_{12}  \delta(\omega^\mathbf{k}_{12})
    \left(
         n_1  n_2
      -  n_\mathbf{k} n_1
      -  n_2 n_\mathbf{k}
    \right),
\end{equation*}
with interacting triads of waves lying on a resonant manifold, enforced by the Dirac delta functions for wavenumbers and frequencies, $\delta^\mathbf{k}_{12} = \delta(\mathbf{k}-\mathbf{k}_1-\mathbf{k}_2)$ and $\delta(\omega^\mathbf{k}_{12}) = \delta(\omega_\mathbf{k}-\omega_1-\omega_2)$ respectively. 
(Note that no special role is played by wave triads that are \textit{exactly} in resonance, as the resonant manifold $\omega_\mathbf{k}=\omega_1+\omega_2$, $\mathbf{k}=\mathbf{k}_1+\mathbf{k}_2$ is broadened by nonlinearity, and encompasses many off-resonant modes, see~\citep{nazarenko2011waveturbbook} for details). The strength of near-resonant interactions is given by the three-wave interaction coefficient
\begin{equation}
\label{eq:Vk12}
	V^\mathbf{k}_{12} 
   = 
	\rho^2\sqrt{\beta  k_x k_{1x} k_{2x}}
	\left( 
		\frac{k_{1y}}{1+ \rho^2 k_1^2}  +  \frac{k_{2y}}{1 + \rho^2 k_2^2}  -  \frac{k_y}{1+\rho^2k^2} 
	\right).
\end{equation} 
In the above we have used the shorthand $n_j=n_{\mathbf{k}_j}$ etc, and likewise for $\omega_j$.

In~\eqref{eq:CHM_kinetic} the integral is taken over half of $\mathbf{k}$-space due to $\psi(\mathbf{x},t)$ being a real field, and hence $\psi_\mathbf{k}^* = \psi_{-\mathbf{k}}$. The zonal axis $k_x=0$ is excluded because by~\eqref{eq:Vk12} these modes do not take part in wave interactions, so modes that are exactly on the zonal axis do not evolve with the rest of the wave spectrum and are not described by~\eqref{eq:CHM_kinetic}.

Details of the derivation of~\eqref{eq:CHM_kinetic}, as well as many of the results obtained by the WT theory for the CHM equation, can be found in review~\citep{ConnNazQuinn2015rossby}.

\subsection{Invariants of the kinetic equation, and the Rayleigh-Jeans spectrum}

The three-wave kinetic equation \eqref{eq:CHM_kinetic} describes the irreversible evolution of an initial spectrum of waves towards states that make the spectrum stationary ($\partial_t n_\mathbf{k}=0$). This kinetic equation has an H-theorem for the growth of nonequilibrium entropy, ensuring ergodic exploration of the phase space in the microcanonical ensemble~\citep{peierls1929kinetischen}.
 As this evolution progresses the dynamical invariants of the equation remain constant, but are redistributed throughout $\mathbf{k}$-space.  We turn our attention to these invariants next.

\subsubsection{Invariants of the kinetic equation}
\label{subsubsec:CHM_invariants}

As in the 2D Euler equation there are two macroscopic invariants---the total enstrophy and energy---that are conserved by the CHM equation~\eqref{eq:CHM}. These two invariants are also exactly conserved by the kinetic equation~\eqref{eq:CHM_kinetic}, and are calculated by integrating the enstrophy and energy spectra $\Omega_\mathbf{k}$ and $E_\mathbf{k}$ respectively, over Fourier space: 
\begin{equation}
	\label{eq:Om_general}
	\Omega    =    \int\limits_{k_x > 0}   \!  \Omega_\mathbf{k}    \, \mathrm{d}\mathbf{k}
		\qquad \qquad \textrm{with} \qquad \qquad
	\Omega_\mathbf{k}    =    k_x n_\mathbf{k},
\end{equation}
and
\begin{equation}
	\label{eq:E_general}
	E    =    \int\limits_{k_x > 0}   \!   E_\mathbf{k}    \, \mathrm{d}\mathbf{k}
		\qquad \qquad \textrm{with} \qquad \qquad
	E_\mathbf{k}    =    \omega_\mathbf{k} n_\mathbf{k},
\end{equation}
i.e.\ $k_x$ and $\omega_\mathbf{k}$ are respectively the enstrophy and energy densities.

In addition to these there is a third independent macroscopic quantity that is conserved by the kinetic equation: 
the zonostrophy~\citep{balknazazakh1990structure, nazarenkoPhDthesis, balknazazakh1991new, balk1991new}
\begin{equation}
	\label{eq:Z_general}
	Z    =    \int\limits_{k_x > 0}  \!  Z_\mathbf{k}    \,\mathrm{d}\mathbf{k} 
		\qquad \qquad \textrm{with} \qquad \qquad
	Z_\mathbf{k}    =    \varphi_\mathbf{k} n_\mathbf{k}.
\end{equation}
Formally $Z$ is only adiabatically conserved by the original CHM equation~\eqref{eq:CHM}. However in simulations it is remarkably well conserved for long times even for strong nonlinearities, see~\citep{nazarenko2009triple}.

In this work we set $\beta=1$ by transforming $(\psi, t) \to (\beta\psi,t/\beta)$ in \eqref{eq:CHM}, and work in the small-scale limit $\rho k \gg 1$, in which case the energy and zonostrophy densities become~\citep{nazarenkoPhDthesis}
\begin{equation} 
	\label{eq:EZ_densities}
	\omega_\mathbf{k} = \frac{k_x}{k^2} 
		\qquad \qquad \textrm{and} \qquad \qquad
	\varphi_\mathbf{k} = \frac{k_x^3(k_x^2 + 5 k_y^2)}{k^{10}}. 
\end{equation}

Importantly, as $k_x > 0$, all three invariants of the kinetic equation $(\Omega, E, Z)$ are positive-definite. Consequently the Fj{\o}rtoft principle described in section~\ref{sec:intro} applies: each invariant places a constraint on where the other two invariants move in $\mathbf{k}$-space during the evolution of the kinetic equation, leading to the dynamical formation of zonal flows~\citep{nazarenko2009triple}. As well as the enstrophy, $\Omega$ also corresponds to the total momentum of the system in the $x$ direction. The $y$-momentum $\int_{k_x > 0} k_y n_\mathbf{k} \mathrm{d}\mathbf{k}$ is also conserved, but it can be positive or negative, and so places no constraint on the flow of the other invariants through $\mathbf{k}$-space. For simplicity in what follows we consider systems with zero $y$-momentum, i.e.\ spectra that are even in $k_y$. We also exclude the meridional axis $k_y=0$ to retain symmetry with the zonal axis and to make the gravest mode in our system the largest-scale isotropic mode, c.f.~\citep{kraichnan1975statistical}. Formally we could retain the meridional axis in what follows, but the gravest mode would then be anisotropic, which would be unrealistic.

    Thus $\mathbf{k}$-space is restricted to the first quadrant $(k_x,k_y) > 0$, and \eqref{eq:Om_general}-\eqref{eq:Z_general} will yeld $(\Omega,E,Z)$ up to a factor of two that can be absorbed into the definition $n_\mathbf{k}$.

\subsubsection{Rayleigh-Jeans equilibrium spectrum}
\label{subsubsec:CHM_RJ}

The amounts of the dynamical invariants that a system of waves possesses are set by the initial spectrum according to~\eqref{eq:Om_general}-\eqref{eq:Z_general}. The spectrum then evolves according to the kinetic equation~\eqref{eq:CHM_kinetic}, which redistributes the invariants across $\mathbf{k}$-space, until thermodynamic equilibrium is reached. Formally the equilibrium solution of~\eqref{eq:CHM_kinetic} is the RJ spectrum
\begin{equation}
	\label{eq:RJspectrum}
	\nRJ = \frac{T}{\mu  \omega_\mathbf{k}  + k_x + \lambda \varphi_\mathbf{k}},
\end{equation}
on which the quantity 
$\sigma = (\mu  E + \Omega + \lambda Z)/T$, 
i.e.\ a linear combination of the three invariants of the kinetic equation, is partitioned equally across the available degrees of freedom (Fourier modes). This can be seen by noting that on the RJ spectrum the density of $\sigma$, i.e. $\sigma_\mathbf{k}$, is independent of $\mathbf{k}$. However, if this the case in an unbounded $\mathbf{k}$-space then the integral of $\sigma_\mathbf{k}$ against spectrum~\eqref{eq:RJspectrum} diverges, which is unphysical. This is the WT version of the ultraviolet catastrophe. To restore nontrivial equilibria we introduce a high wavenumber cutoff $k_\mathrm{max}$, corresponding to a minimum allowed lengthscale in the system. In a real fluid or plasma this could represent the effects of small-scale physics; in simulations this is the spatial discretisation~\citep{connaughton2005condensation}.

In the RJ spectrum~\eqref{eq:RJspectrum} we refer to $T$ as the ``temperature'', $\mu$ as the ``chemical potential'' and $\lambda$ as the ``zonostrophy potential''. This  differs from the usual WT convention in which $1/T$  is conjugate to the energy density $\omega_\mathbf{k}$.  In~\eqref{eq:RJspectrum} we have instead selected $1/T$ to be conjugate to the enstrophy density $k_x$ to make contact with both the large-scale condensation of energy in the 2D Euler equation, and particles in the 2D GPE. The equilibrium energy spectrum for the 2D Euler equation has the same funcitonal form as the waveaction spectrum for the 2D GPE~\citep{kraichnan1975statistical, kraichmonty1980}. These coincide with the CHM energy spectrum $\ERJ = \omega_\mathbf{k}\nRJ$ when we set $\lambda=0$. Likewise the CHM energy and enstrophy invariants with $\lambda=0$ have the same functional form as those of the 2D Euler equation, and the total waveaction and energy in the 2D GPE, see table~\ref{tab:CHM_GPE}.
	Thus when we set $\lambda=0$, the potentials  $\mu$ and $T$ in \eqref{eq:RJspectrum} directly map on to the chemical potential and temperature of an analogous GPE system.
	
\begin{table}[t]
\caption{\label{tab:CHM_GPE}Equivalence between equilibrium spectra and invariants of the CHM equation, GPE, and 2D Euler equation.}
\begin{indented}
\item[]
\begin{tabular}{@{}llll}
\br
CHM ($\lambda=0$)                 & 2D Euler                  & GPE                           &  Functional form \  \\
\hline 
$\ERJ$ 									 & $\ERJ$					 & $\nRJ$    					&  $\frac{T}{k^2+\mu}$\\[1mm]
$E$    						 	          &  $E$                       &  $\mathcal{N}$				 & $\int \! \frac{T}{k^2+\mu}\, \mathrm{d}\mathbf{k}$\\[1mm]
$\Omega$ 	                            &  $\Omega$             &  $\mathcal{E}$                & $\int \! \frac{Tk^2 }{k^2+\mu}\, \mathrm{d}\mathbf{k}$\\
\br
\end{tabular}
\end{indented}
\end{table}

At this point we re-introduce the finite box size $L$, in order to study the behaviour of the RJ spectrum and invariants as $L\to\infty$. This involves rewriting the integrals in~\eqref{eq:CHM_kinetic}-\eqref{eq:Z_general} as discrete sums over a lattice of modes in Fourier space, with mode spacing $\Delta k\!=\!2\pi/L$, from $k_\mathrm{min}\!=\!\Delta k$ to $k_\mathrm{max}$ in each direction.
We refer to this as the ``discrete sum'' approach. (This is distinct from ``discrete wave turbulence'' where no kinetic equation can be derived; see~\citep{harper2013quadratic} for a discussion of the latter in the CHM equation.)
We note, however, that the derivation of the kinetic equation~\eqref{eq:CHM_kinetic} relies on taking the $L\to\infty$. Here we assume without proof that the conservation of $(\Omega,E,Z)$ remains approximately valid for a large but finite system containing many Fourier modes, although for illustration purposes our figures will be plotted with only a few modes for clarity.

Finally, we can eliminate one free parameter by rescaling variables in terms of $k_\mathrm{max}$:
\begin{equation}
\label{eq:CHM_rescale_vbles}
	\epsilon					= \frac{k_\mathrm{min}}{k_\mathrm{max}},
		\qquad   
	\tilde{\mathbf{k}}	= \frac{\mathbf{k}}{k_\mathrm{max}},
		\qquad 
	\tilde{\mu}				= \frac{\mu}{k_\mathrm{max}^2},
		\qquad 
	\tilde{\lambda}		= \frac{\lambda}{k_\mathrm{max}^6},
		\qquad
	\tilde{\Omega}		= \frac{\Omega}{k_\mathrm{max}^2},
		\qquad 
	\tilde{Z}					= k_\mathrm{max}^4 Z,
\end{equation}
($E$ is unchanged by this rescaling).
Dropping the tildes immediately, the quantity of the three invariants that can be fit by a Rayleigh-Jeans equilibrium spectrum are given by 
\begin{equation}
	\label{eq:invariants_RJ_discr}
	\Omega 	=  T\epsilon^2 \sum_{L_\epsilon} \! \frac{k^{10}}{D(\mathbf{k},\mu,\lambda)},  
\qquad
	E				=  T\epsilon^2 \sum_{L_\epsilon} \! \frac{k^8}{D(\mathbf{k},\mu,\lambda)},  
\qquad
	Z				=  T\epsilon^2 \sum_{L_\epsilon} \! \frac{k_x^2 (k_x^2 + 5k_y^2)}
                             {D(\mathbf{k},\mu,\lambda)},  
\end{equation}
when we are using discrete sums. All three summands have denominator
\begin{equation*}
	D(\mathbf{k},\mu,\lambda)=\mu k^8 + k^{10} + \lambda k_x^2 (k_x^2 + 5k_y^2),
\end{equation*}
and the sums are over the $\mathbf{k}$-space lattice $L_\epsilon = \{(k_x,k_y) = (\epsilon i, \epsilon j): i,j=1,\ldots,\epsilon^{-1}\}$.

We use discrete sums in order to consistently study the infinite box limit $\epsilon\to 0$ in which both the mode spacing and the lower limits of the integrals in~\eqref{eq:Om_general}-\eqref{eq:Z_general} vanish together. If instead we let send the mode spacing to zero while retaining a finite lower limit on the integrals then the integration would be over the square $S_\epsilon=[\epsilon,1]^2$. We will refer to this in passing as the ``continuous integral'' approach, in contrast to the discrete sum approach.


\section{Regularity and divergence of the RJ spectrum}
\label{sec:RJdiverg}

Having established the form of the RJ spectrum, we can now explore the parameter range for which it is valid. The waveaction spectrum~\eqref{eq:CHM_nk} is manifestly positive-definite in $k_x>0$, and so we require $\sgn T = \sgn D(\mathbf{k}, \mu, \lambda)$. Therefore physically meaningful equilibria exist with positive and negative temperature, as long as the spectrum, and hence the invariants~\eqref{eq:invariants_RJ_discr}, are positive.
On the other hand, the spectrum diverges, and the RJ solution is no longer physical, when 
\begin{equation}
	\label{eq:Deq0}
D(\mathbf{k},\mu,\lambda)=0.
\end{equation}

In this section we  establish the conditions for~\eqref{eq:Deq0} to be true and hence map out the edge of parameter space corresponding to allowable RJ spectra. The set corresponding to~\eqref{eq:Deq0} is some 3D hypersurface in the space of $(k_x, k_y,\mu,\lambda)$. We can collapse this to a 2D plot by expressing one of these variables as a function of one other, treating the other two as parameters. In section~\ref{subsec:lambdamu} we (arbitrarily) choose $\lambda$ as the independent variable and plot $\mu(\lambda)$, treating $k_x$ and $k_y$ as parameters.
Then in section~\ref{subsec:kxky} we plot the $k_x$--$k_y$ plane, with parameters $\lambda$ and $\mu(\lambda)$ set to their values that first make the RJ spectrum singular.

\subsection{The $\lambda$---$\mu$ plane; $\mu^+_-\!(\lambda)$ are the boundaries of the RJ-accessible region}
\label{subsec:lambdamu}

\begin{figure}
	\centering 
	\includegraphics[width=0.999\textwidth]{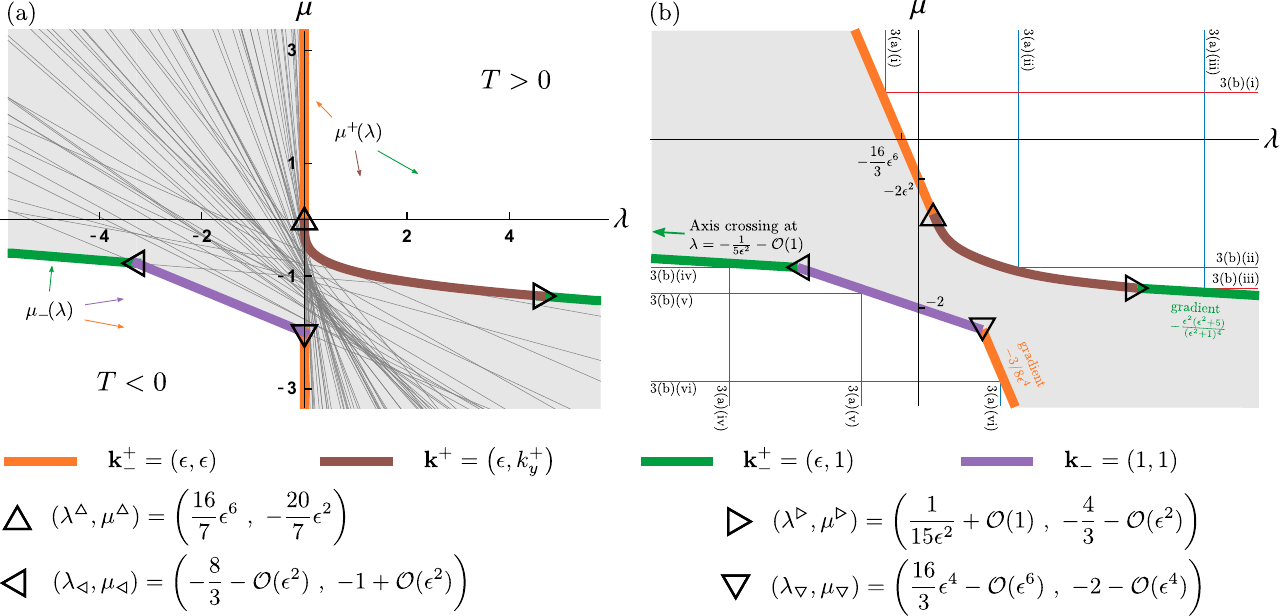}
	\caption{The $\lambda$--$\mu$ plane, with regions where the RJ spectrum is accessible (white) 
				separated from the RJ-forbidden (grey) region by the RJ boundary comprised of the coloured curves 
				$\mu^+\!(\lambda)$  and $\mu_-\!(\lambda)$.
				(a) Accurate plot of $\lambda$--$\mu$ plane with $\epsilon=1/8$. 
				(b) Schematic sketch of the $\lambda$--$\mu$ plane. Labels on lines of constant  $\lambda$ (vertical, blue) 
				and $\mu$ (horizontal, red) correspond to panels of figure~\ref{fig:Condensation_convergence}.
				}
	\label{fig:mulambda}
\end{figure}

First, we solve~\eqref{eq:Deq0} for $\mu$. This gives a set of straight lines in the $\lambda$--$\mu$ plane,
\begin{equation}
	\label{eq:mulines}
	\mu(\mathbf{k},\lambda) = -k^2 -  \frac{k_x^2(k_x^2 + 5k_y^2)}{k^{8}} \lambda,
\end{equation}
on which the RJ spectrum diverges: one line for each Fourier mode $\mathbf{k}\in L_\epsilon$. These lines are drawn in dark grey in figure~\ref{fig:mulambda}(a) for $\epsilon=1/8$.  (If we use continuous integrals then there is one line for every $\mathbf{k}\in S_\epsilon$, and the lines densely sweep out a ruled surface, shown in light grey in figure~\ref{fig:mulambda}(a) and~\ref{fig:mulambda}(b).)

This divides the $\lambda$--$\mu$ plane into three portions: above (white in figure~\ref{fig:mulambda}) are $(\lambda, \mu)$ combinations that give physical RJ states with $T>0$;  below (also white) are RJ states with $T<0$; and the region in between (grey) are $\lambda,\mu$ combinations for which $n_\mathbf{k}$ diverges for at least one $\mathbf{k}\in L_\epsilon$ (or $\mathbf{k}\in S_\epsilon$ for continuous integrals). As we will see in section~\ref{subsec:negativepotentials}, for $(\lambda,\mu)$ points within the grey region there are some modes $\mathbf{k}$ for which $\nRJ$ is negative, which is unphysical. The grey region thus contains the parameter values which are forbidden for a physical RJ spectrum; we will sometimes refer to it as the ``RJ-forbidden region''. For $(\lambda,\mu)$ points exactly on the boundary of the grey region (the ``RJ boundary'' in what follows), $\nRJ$ diverges at one or two modes, and these will represent condensates.

The upper and lower boundaries between the white RJ-accessible regions and the grey RJ-forbidden region are the piecewise linear functions\footnote{
		We use superscripts on variables when referring to 
		their values on the upper RJ boundary, i.e.\ with positive $T$, 
		subscripts to refer to their values on the negative $T$ boundary, and both scripts 
		simultaneously, e.g.~$\mu^+_-\!(\lambda)$, $\mathbf{k}^+_-\!(\lambda)$ when speaking of both boundaries 
		at once. Had we chosen $\mu$ as the independent variable we could define $\lambda^+_-\!(\lambda)$ and 
		$\mathbf{k}^+_-\!(\mu)$ analogously. For compactness we sometimes suppress explicit dependence on the independent variable, 
		when the context is clear. 
		}
\begin{equation}
	\label{eq:muPlusMinus}
	\mu^+\!(\lambda) = \max_{\mathbf{k}\in L_\epsilon} \mu(\mathbf{k},\lambda)
		\qquad \mathrm{and} \qquad
	\mu_-\!(\lambda) = \min_{\mathbf{k}\in L_\epsilon} \mu(\mathbf{k},\lambda),
\end{equation}
respectively (in the continuous integral approach the $\max$ and $\min$ are taken over $\mathbf{k}\in S_\epsilon$ and the brown portion of of $\mu^+\!(\lambda)$ is a differentiable function, see section~\ref{subsec:kxky}). 
	Both the $\mu^+\!(\lambda)$ and $\mu_-\!(\lambda)$ RJ boundaries divide into three segments. In each segment a different mode $\mathbf{k}^+\!(\lambda)$ or $\mathbf{k}_-\!(\lambda)$ parameterises that part of the boundary, i.e.\
\begin{equation}
	\label{eq:kPlusMinus}
	\mathbf{k}^+\!(\lambda) = \argmax_{\mathbf{k}\in L_\epsilon} \mu(\mathbf{k},\lambda)
		\qquad \mathrm{or} \qquad
	\mathbf{k}_-\!(\lambda) = \argmin_{\mathbf{k}\in L_\epsilon} \mu(\mathbf{k},\lambda)
\end{equation}
(or $\argmax$ and $\argmin$ over $\mathbf{k}\in S_\epsilon$ for continuous integrals). In figure~\ref{fig:mulambda} the different segments of $\mu^+_-$ are coloured according to the $\mathbf{k}^+_-$ that parameterises them; we retain the colour scheme throughout this paper to refer to the various $\mathbf{k}^+_-$.

The points where the mode $\mathbf{k}^+_-$ changes are marked by $\medtriangleup$, $\medtriangleright$, $\medtriangledown$, $\medtriangleleft$. Their $(\lambda,\mu)$ coordinates are given in the key to leading order in $\epsilon$. Next, we display the $\mathbf{k}^+_-$ in the $k_x$--$k_y$ plane and identify their values.

\subsection{The $k_x$--$k_y$ plane: $\mathbf{k}^+_-$ are the modes at which $\nRJ$ becomes singular}
\label{subsec:kxky}

\begin{figure}
	\centering 
		\includegraphics[width=\textwidth]{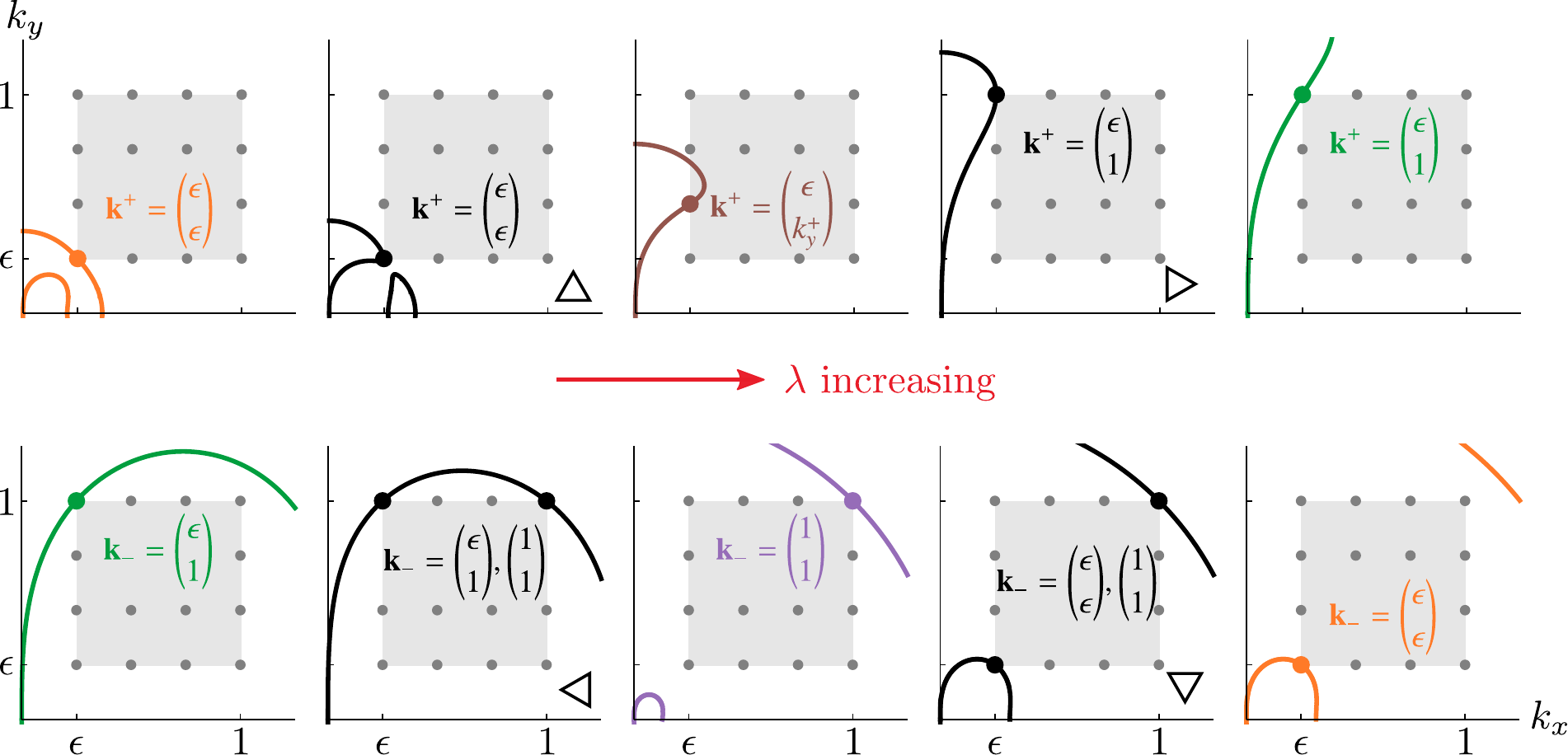}
	\caption{$k_x$--$k_y$ plane showing the curve $D=0$ intersecting $L_{1/4}$ (dark grey lattice) 
				at the mode $\mathbf{k}^+$ (top row, for the boundary of the $T>0$ region) 
				or $\mathbf{k}_-$ (bottom row, for the boundary of the $T<0$ region).
					Typical configurations are shown for each regime of $\mathbf{k}^+_-\!(\lambda)$ detailed in~\eqref{eq:kPlus}
				and~\eqref{eq:kMinus}, distinguished by colour.
					Panels labelled with black triangles correspond to the transitions between 
				the different $\mathbf{k}^+_-\!(\lambda)$ regimes. The light grey square is $S_{1/4}$.
				}
	\label{fig:kxky}
\end{figure}

In figure~\ref{fig:kxky}, each panel shows the $k_x$--$k_y$ plane, with the lattice $L_\epsilon$ marked in dark grey points and the square $S_\epsilon$ in light grey. We plot the parametric curves $D(\mathbf{k},\mu^+\!(\lambda),\lambda)=0$ (top row) and  $D(\mathbf{k},\mu_-\!(\lambda),\lambda)=0$ (bottom row), with $\lambda$ increasing from left to right. 

In each panel $\mathbf{k}^+_-\!(\lambda)$ is shown by the coloured dot, and is the wavenumber where the $D=0$ curve first intersects a point in $L_\epsilon$ (or first touches $S_\epsilon$ in the continuous integral approach). Colours for $\mathbf{k}^+_-\!(\lambda)$  and the $D=0$ curve correspond to the coloured portions of $\mu^+_-\!(\lambda)$ in figure~\ref{fig:mulambda}, and label the mode $\mathbf{k}^+_-\!(\lambda)$ for the ranges of $\lambda$ we now detail.

\subsubsection{Bottom row of figure~\ref{fig:kxky}.}

The $T<0$ case is simpler so we consider it first to fix concepts. Expressions for $\mathbf{k}_-\!(\lambda)$ can be found by minimising $\mu(\mathbf{k} , \lambda)$ on $L_\epsilon$ by elementary means (identical results are found when we use continuous integrals). We obtain
\begin{equation}
\label{eq:kMinus}
\mathbf{k}_-\!(\lambda) =
	\begin{cases}
		\binom{\epsilon}{1}				& \mbox{for } \lambda \leq \lambda_\medtriangleleft = - \frac{8(1+\epsilon^2)^4}{3-25\epsilon^2-15\epsilon^4-3\epsilon^6},\\
		\binom{1}{1} 						& \mbox{for } \lambda_\medtriangleleft \leq \lambda \leq \lambda_\medtriangledown,\\
		\binom{\epsilon}{\epsilon}  & \mbox{for } \lambda \geq \lambda_\medtriangledown=\frac{16}{3}\frac{\epsilon^4}{1+\epsilon^2}.	
	\end{cases}
\end{equation}
Each panel of figure~\ref{fig:kxky} shows a representative configuration for the $D=0$ curve, and the corresponding location for $\mathbf{k}_-$, as one moves from left to right along the bottom RJ boundary $\mu_-\!(\lambda)$ in figure~\ref{fig:mulambda}. The mode $\mathbf{k}_-$ over which $\nRJ$ diverges is at first at the smallest zonal scale; then isotropic at the smallest scale of the system; then isotropic at the largest scale.

Transitioning between these three cases (we we will refer to the three cases in~\eqref{eq:kMinus} as ``$\mathbf{k}_-$ regimes'' in what follows), $\mathbf{k}_-$ jumps discontinuously, due to the $D=0$ curve touching $L_\epsilon$ at two different Fourier modes simultaneously. These transitions are shown in black in figure~\ref{fig:kxky}, in the panels labelled $\medtriangleleft$ and $\medtriangledown$. They are marked likewise in the $\lambda$--$\mu$ plane in figure~\ref{fig:mulambda}, and occur at coordinates $(\lambda_\medtriangleleft,\mu_\medtriangleleft)$ and $(\lambda_\medtriangledown,\mu_\medtriangledown)$, whose leading order behaviour with $\epsilon$ is given in the key. 

\subsubsection{Top row of figure~\ref{fig:kxky}.}
\label{subsubsec:kxkyTpos}

For the $T>0$ case we maximise $\mu(\mathbf{k},\lambda)$ to find 
\begin{equation}
\label{eq:kPlus}
\mathbf{k}^+\!(\lambda) =
	\begin{cases}
		\binom{\epsilon}{\epsilon}					&\mbox{for } \lambda \leq \lambda^\medtriangleup = \frac{16\epsilon^6}{7},\\
		\binom{\epsilon}{k_y^+\!(\lambda)}	&\mbox{for } \lambda^\medtriangleup \leq \lambda \leq \lambda^\medtriangleright,\\
		\binom{ \epsilon}{1}							&\mbox{for } \lambda \geq \lambda^\medtriangleright =\frac{(1+\epsilon)^5}{\epsilon^2(15-\epsilon^2)}.
	\end{cases}
\end{equation}
Between these three cases (``henceforth $\mathbf{k}^+$ regimes''), $\mathbf{k}^+\!(\lambda)$ transitions with no jump, see the panels labelled $\medtriangleup$ and $\medtriangleright$ in figure~\ref{fig:kxky}. The transitions are marked likewise in figure~\ref{fig:mulambda}, with $(\lambda,\mu)$ coordinates $(\lambda^\medtriangleup,\mu^\medtriangleup)$ and $(\lambda^\medtriangleright,\mu^\medtriangleright)$, given to leading order in $\epsilon$ in the key.

In~\eqref{eq:kPlus}, $k_y^+\!(\lambda)$ is different when discrete sums are used in~\eqref{eq:invariants_RJ_discr} vs.\ using continuous integrals. 
In the latter, $\nRJ$ becomes divergent when the $D=0$ curve first meets the square $S_\epsilon$, i.e. $(\epsilon,k_y^+)$ is the point where $S_\epsilon$ is tangent to $D=0$. Correspondingly, $k_y^+$ is the real root in $[\epsilon,1]$ of the 10th-degree polynomial found by solving $\partial_{k_y}\mu(k_x\!\!=\!\!\epsilon, k_y, \lambda)=0$, and increases continuously with $\lambda$.

When discrete sums are used the $D=0$ curve is only constrained by the lattice $L_\epsilon$, and can penetrate between the lattice points slightly, as exemplified in the middle panel, top row of figure~\ref{fig:kxky}. Thus, $k_y^+=j\epsilon$ for $j=1,\ldots,\epsilon^{-1}$, and increases piecewise with $\lambda$ so that $\mathbf{k}^+=(\epsilon, k_y^+)$ transitions discontinuously up the left-hand edge of $L_\epsilon$. At the transitions the $D=0$ curve touches two adjacent lattice points and $\mathbf{k}^+$ represents two neighbouring Fourier modes.

For $T>0$, then, moving along the top RJ boundary $\mu^+\!(\lambda)$ in figure~\ref{fig:mulambda}, the mode $\mathbf{k}^+$ over which $n_\mathbf{k}$ diverges is of the largest scale in both $x$ and $y$  (fully isotropic); remains largest in $x$ and shrinks in $y$ (becoming more zonally anisotropic as $\lambda$ increases); and finally ends up at the largest possible scale in the $x$-direction and smallest scale in the $y$-direction (smallest-scale zonal mode). 

For clarity we have set $\epsilon=1/4$ in figure~\ref{fig:kxky}; the behaviour is qualitatively the same for smaller values of $\epsilon$. In each panel of figure~\ref{fig:kxky} we have displayed representatives of each $\mathbf{k}^+_-\!(\lambda)$ regime detailed in \eqref{eq:kMinus}, \eqref{eq:kPlus}. Within each regime the $\mathbf{k}^+_-\!(\lambda)$ remain the same despite the $D=0$ curve changing shape, or even topology, as $\lambda$ and $\mu^+_-\!(\lambda)$ change.

\subsection{Regularity of the RJ spectrum, negative thermodynamic potentials}	
\label{subsec:negativepotentials}

In summary, every $(\lambda ,\mu)$ point lying in either of the white ($T>0$ or $T<0$) regions in figure~\ref{fig:mulambda} gives  summands over $L_\epsilon$ in~\eqref{eq:invariants_RJ_discr} that are all positive, corresponding to a physical RJ spectrum. In figure~\ref{fig:mulambda}(b) we sketch the $\lambda$--$\mu$ plane schematically, with the axes intercepts labelled, to show that RJ states are possible for any sign combination of $(T,\mu,\lambda)$, except for the case $T<0, \mu>0, \lambda>0$. We also show (vertical blue) lines of constant $\lambda$ and (horizontal red) lines of constant $\mu$ in the RJ-accessible regions. These schematically show the lines of approach to the RJ boundary corresponding to the panels of figure~\ref{fig:Condensation_convergence}, and labelled accordingly.

As we will discuss in the next section, when the $(\lambda,\mu)$ point approaches the RJ boundary, the spectrum $\nRJ$ becomes increasingly peaked at the $\mathbf{k}^+_-$ corresponding to that boundary point. Thus the various $\mathbf{k}^+_-$ are the fundamental modes into which the invariants will condense on the RJ spectrum. In the CHM  the fundamental mode of condensation shows great diversity---isotropic, anisotropic, small-scale and large-scale---as compared to the isotropic 2D GPE or Euler equations, for which the condensation occurs only at either the gravest or smallest-scale isotropic modes. 

Inspection of figure~\ref{fig:mulambda}(b) shows that proximity to the boundary, and hence condensation, is associated with at least one thermodynamic potential becoming negative. However this is not a sufficient condition for condensation as there are many regions of the $\lambda$--$\mu$ plane that have negative potentials but are remote from the RJ boundary, for example deep inside the quadrant $(T,\mu,\lambda)\ll1$. 
	Thus, while it is necessary for at least one of $(T,\mu,\lambda)$ to be negative in order to have a condense, this condition is not sufficient. In the next section we will see that the necessary proximity of the $(\lambda,\mu)$ point to the RJ boundary in order to have condensation depends strongly on $\epsilon$.

Finally, as the $(\lambda, \mu)$ point passes inside the RJ-forbidden grey region in $\lambda$--$\mu$ space, the $D=0$ curve in $k_x$--$k_y$ space envelops at least one lattice point $\mathbf{k}\in L_\epsilon$. This means that $D$, and hence $\nRJ$, changes sign for that $\mathbf{k}$, so the RJ spectrum is no longer positive-definite and ceases to remain physically meaningful. If we are using continuous integrals, when the $(\lambda,\mu)$ point passes inside the RJ-forbidden region, the $D=0$ curve in $k_x$--$k_y$ space penetrates $S_\epsilon$ and envelops a portion of $S_\epsilon$ with finite area. At those ``captured'' modes $\nRJ$ will become negative, which is unphysical. As discussed in section~\ref{subsubsec:kxkyTpos}, the $D=0$ curve can penetrate between lattice points of $L_\epsilon$ slightly, but this captures a finite-area portion of $S_\epsilon$. Thus the grey RJ-forbidden region in $\lambda$--$\mu$ space is thus slightly bigger when using continuous integrals vs.\ using discrete sums.

\section{Condensation in the RJ spectrum}
\label{sec:condensation}

Consider a sequence of equilibrium states lying on a vertical blue line of constant $\lambda$ in figure~\ref{fig:mulambda}(b). Starting from a point deep in the RJ-accessible region and approaching the boundary $\mu^+_-\!(\lambda)$, the curve $D=0$ approaches the $L_\epsilon$ lattice, and the spectrum becomes increasingly peaked over the mode(s) $\mathbf{k}^+_-\!(\lambda)$. Finally when  $\mu=\mu^+_-\!(\lambda)$, the curve touches $L_\epsilon$ at $\mathbf{k}^+_-$, as typified by one of the panels in figure~\ref{fig:kxky}, and the spectrum $\nRJ$ becomes singular at $\mathbf{k}^+_-$. 	Likewise we could approach the RJ boundary along a horizontal red line in figure~\ref{fig:mulambda}(b), keeping $\mu$ fixed and approaching $\lambda^+_-\!(\mu)$, with the spectrum becoming increasingly peaked, and finally singular, at $\mathbf{k}^+_-\!(\mu)$. 
		
As the spectrum at $\mathbf{k}^+_-$ becomes increasingly peaked, the term (or terms, in the cases where $\mathbf{k}^+_-$ represents two different Fourier modes) containing $\mathbf{k}^+_-$ in~\eqref{eq:invariants_RJ_discr} starts to dominate the sum. Thus, as $(\lambda,\mu)$ moves to the edge of the RJ-accessible region, the spectrum becomes a Kronecker delta (equal to unity when its argument $\mathbf{k}-\mathbf{k'}=0$ and zero otherwise):
\begin{equation}
	\label{eq:nRJdelta}
	\nRJ \to \delta(\mathbf{k} - \mathbf{k}^+_-),
\end{equation}
or a linear combination of  Kronecker deltas, when $\mathbf{k}^+_-$ represents two different modes,
\begin{equation}
	\label{eq:nRJtwodeltas}
	\nRJ \to \alpha \, \delta(\mathbf{k} - \mathbf{k}_1) + (1-\alpha) \, \delta(\mathbf{k} - \mathbf{k}_2),
	\qquad 
	0 < \alpha < 1,
\end{equation}
for example $\mathbf{k}_1=(\epsilon,1)$, $\mathbf{k}_2=(1,1)$ at the $\medtriangleleft$ point.

Eventually in the singular limit~\eqref{eq:nRJdelta} (or~\eqref{eq:nRJtwodeltas}) all three invariants are contained in the mode(s) $\mathbf{k}^+_-$ while the rest of $\mathbf{k}$-space contributes a vanishingly small amount: the invariants have condensed at the fundamental mode(s) $\mathbf{k}^+_-$. Simultaneously we must have $T\to0$ in order to keep the total $(\Omega, E, Z)$ constant. We stress that this condensation happens entirely within the RJ spectrum, unlike the case for the 3D GPE where a singular spectrum at $\mathbf{k}=0$ arises to absorb the excess particles below a finite transition temperature.

\begin{figure}
	\centering 
	\includegraphics[width=0.999\textwidth]{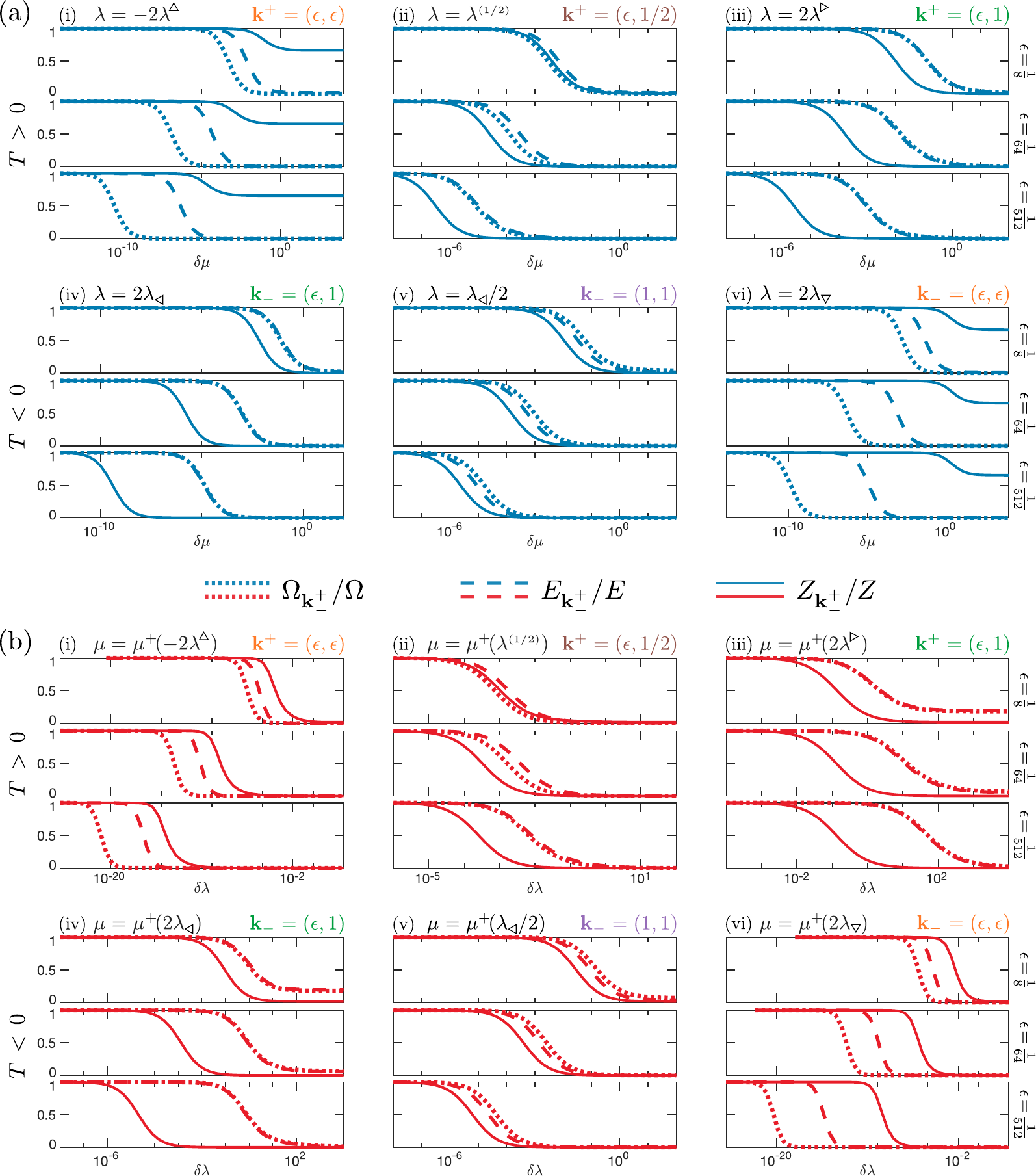}
	\caption{Fraction of the total invariants that are contained within mode $\mathbf{k}^+_-$ as a function of (a) $\delta\mu$, or (b) $\delta\lambda$,
				for the representative points on the RJ boundary within each regime of $\mathbf{k}^+_-$ shown schematically in figure~\ref{fig:mulambda}(b).
				$T>0$ for the top row in parts (a) and (b) and $T<0$ in the bottom row in both parts.
				Within each triptych, each panel represents a different system size $\epsilon$. 
				}
	\label{fig:Condensation_convergence}
\end{figure}

We examine the approach to condensation in figure~\ref{fig:Condensation_convergence}, where in each panel of three triptychs we plot the ratio of each invariant that is contained in mode $\mathbf{k}^+_-\!(\lambda)$ to the total, for example $\Omega_{\mathbf{k}^+_-}/\Omega$ etc.,\ as we approach points on the RJ boundary representative of each $\mathbf{k}^+_-$. These lines of approach are shown schematically in figure~\ref{fig:mulambda}(b). 
 
 	In figure~\ref{fig:Condensation_convergence}(a) we plot these fractions as a function of $\delta\mu=|\mu-\mu^+_-\!(\lambda)|$, i.e.\ moving along a blue line in figure~\ref{fig:mulambda}(b) towards $\mu^+_-\!(\lambda)$. $T>0$ in the top row of triptychs  and $T<0$ in the bottom row. (In panel~\ref{fig:Condensation_convergence}(a,b)(ii), $\lambda^{(1/2)}$ is the value that makes $k_y^+\!(\lambda^{(1/2)})=1/2$.) 

In figure~\ref{fig:Condensation_convergence}(b) we approach the same points on the RJ boundary horizontally, along the red lines in figure~\ref{fig:mulambda}(b), plotting the fractions of invariants contained in mode $\mathbf{k}^+_-$ as a function of $\delta\lambda=|\lambda-\lambda^+_-\!(\mu)|$; again $T>0$ in the top row and $T<0$ in the bottom row.

Within each triptych, the system size increases moving down the panels. 
The plots are qualitatively  similar whichever locations on the RJ boundary are chosen within the $\mathbf{k}^+_-$ regimes.

 \subsection{Finite-size condensation at the edge of the RJ region}
\label{subsec:condensation_dominance}

\begin{figure}[t]
	\centering
	\includegraphics[width=8.6cm]{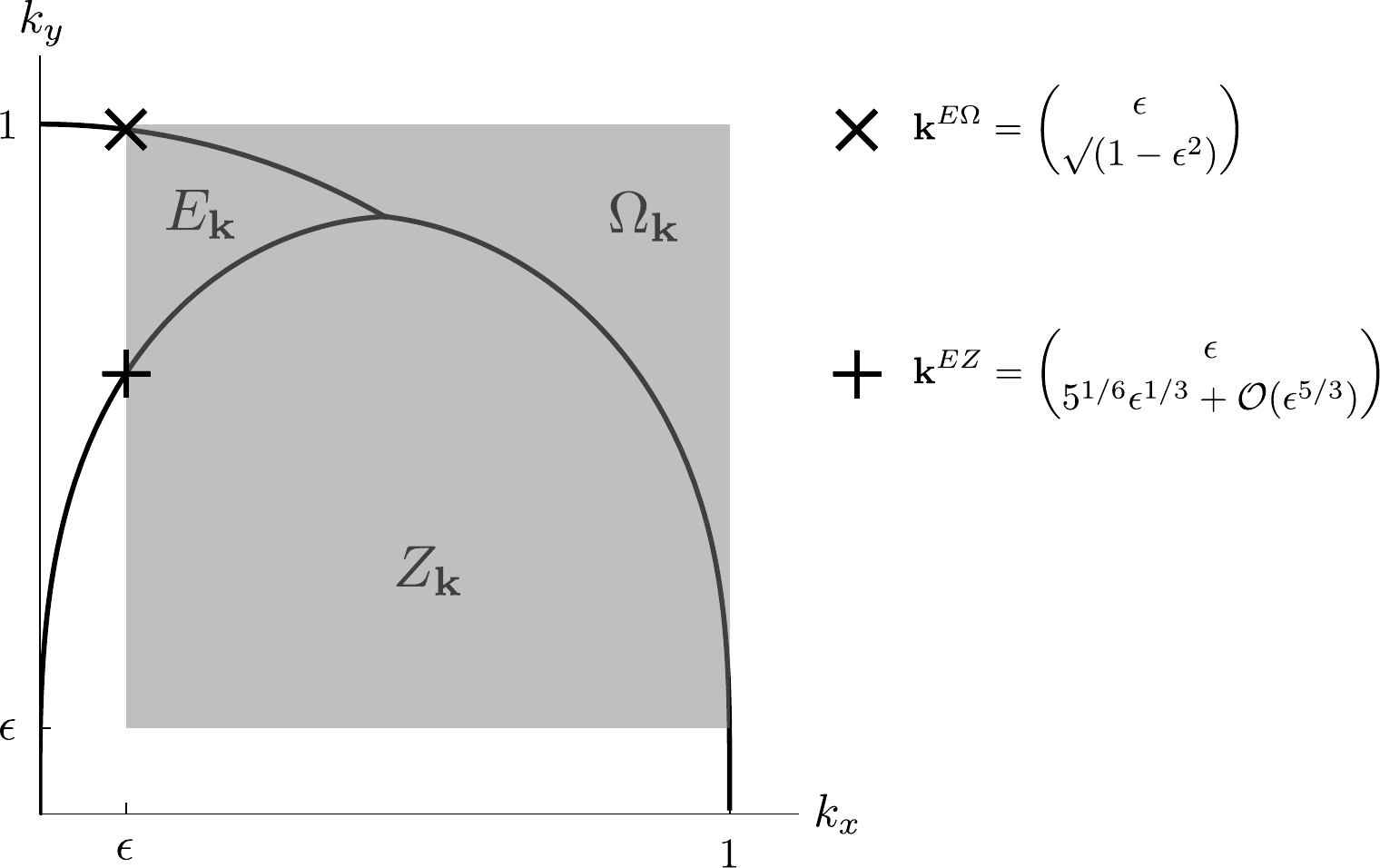}
	\caption{
				Sectors of $\mathbf{k}$-space labelled by the invariant spectrum that is dominant deep inside each sector.
				The grey square is $S_{1/8}$, whose left-hand edge intersects the isolines $E_\mathbf{k}\!=\!\Omega_\mathbf{k}$ 
				and $E_\mathbf{k}\!=\!Z_\mathbf{k}$ at the points marked by $\bigtimes$ and $\bigplus$ respectively.
				}
	\label{fig:InvDensityIsolines}
\end{figure}

Examining figure~\ref{fig:Condensation_convergence}, we see that as we approach the RJ boundary from either direction, all three invariants  condense into the fundamental mode $\mathbf{k}^+_-$ for any finite $\epsilon$, as predicted above. We also see that for different parts of the RJ boundary (different $\mathbf{k}^+_-$ regimes), different invariants condense first, i.e.\ at larger $\delta\lambda$ or $\delta\mu$. 
Which invariant condenses first can be explained by considering which invariant spectrum is dominant at the $\mathbf{k}^+_-$ under consideration.
In figure~\ref{fig:InvDensityIsolines} we plot the isolines where each pair of invariant spectra are equal ($E_\mathbf{k}=Z_\mathbf{k}$ etc.). This divides $L_\epsilon$ into three sectors, deep inside of which one invariant will dominate the other two. The strength of this dominance at a particular $\mathbf{k}$ goes as the ratio of the dominant invariant to the others, at the $\mathbf{k}$ in question.\footnote{
	As mentioned in section~\ref{sec:intro}, figure~\ref{fig:InvDensityIsolines} is also instrumental 
	in applying the Fj{\o}rtoft argument to weakly nonlinear CHM systems. 
	If the system is forced at a particular $\mathbf{k}_\mathrm{f}$ then one can draw three sectors 
	whose boundaries intersect at $\mathbf{k}_\mathrm{f}$ exactly as in figure~\ref{fig:InvDensityIsolines}. 
	Each sector will host the cascade of the invariant whose density dominates there, see~\citep{nazarenko2009triple}. 
}

Thus, when $\nRJ$ is becoming increasingly concentrated at $\mathbf{k}^+_-=(\epsilon,\epsilon)$, it is $Z_\mathbf{k}$ that dominates there and so zonostrophy condenses first into that mode, as we find in figure~\ref{fig:Condensation_convergence}(a,b)(i) and  \ref{fig:Condensation_convergence}(a,b)(vi). Note that the $\bigplus$ point on the $E_\mathbf{k}=Z_\mathbf{k}$ boundary converges to the origin as  $\epsilon^{1/3}$. This is slower than the convergence of the mode $(\epsilon,\epsilon)$, guaranteeing that this mode remains in the $Z_\mathbf{k}$ sector as $\epsilon\to0$, and so is always associated with zonostrophy condensing first.

Similar arguments show that the concentration of $\nRJ$ into zonal modes $\mathbf{k}^+=(\epsilon,k_y^+)$ is associated with energy condensing first.
In figure~\ref{fig:Condensation_convergence}(a,b)(ii) we monitor condensation into $\mathbf{k}^+=(\epsilon,1/2)$. The ratio of energy to enstrophy densities is $E_\mathbf{k}/\Omega_\mathbf{k} = k^{-2}$, which goes as $\sim 1/4$ at this mode as $\epsilon\to 0$, so $E$ condenses ahead of $\Omega$ by a factor which tends to a constant. By contrast $E_\mathbf{k}/Z_\mathbf{k} = k^8/k_x^2$ so the condensation of $Z$ lags that of $E$ at a rate $\sim \! \epsilon^{-2}$ at this mode.

When $\nRJ$ becomes concentrated into the smallest-scale zonal mode $\mathbf{k}^+_-=(\epsilon,1)$, this mode lies near the boundary of the  $\Omega_\mathbf{k}$ and $E_\mathbf{k}$ sectors (the $\bigtimes$ point on this boundary converges quickly to the point $(\epsilon,1)$ so as $\epsilon\to0$), hence the simultaneous condensation of enstrophy and energy into this mode seen in figure~\ref{fig:Condensation_convergence}(a,b)(iii), and \ref{fig:Condensation_convergence}(a,b)(iv). 

Finally for concentration of $\nRJ$ into $\mathbf{k}_-=(1,1)$, this point lies deep in the $\Omega_\mathbf{k}$ sector, hence enstrophy condensing first in figures~\ref{fig:Condensation_convergence}(a,b)(v).

To summarise, for any system of finite size there exists a layer in $\lambda$--$\mu$ space adjacent to the RJ boundary, within which we expect all three invariants to condense into mode $\mathbf{k}^+_-$. Although this is important to note, we do not expect these states to be dynamically significant as the only relevant initial conditions that will equilibrate to these states have spectra that are already highly concentrated around $\mathbf{k}^+_-$. This is to say that the very edge of the RJ region where all invariants condense will be accessed by initial conditions lying in an insignificant volume in the function space of all possible initial conditions.

However, a wider ``boundary layer'' of $(\lambda,\mu)$ values has RJ spectra that manifest condensates of one invariant, if the corresponding $\mathbf{k}^+_-$ lies deep inside the sector in figure~\ref{fig:InvDensityIsolines} in which that invariant density dominates. (If the $\mathbf{k}^+_-$ lies near the boundary between two sectors then the wider boundary layer represents condensates of those two invariants, e.g.\ condensation of $\Omega$ and $E$ but not $Z$ for $\mathbf{k}^+_-=(\epsilon,1)$.) This boundary layer of $(\lambda,\mu)$ values corresponds to the set in the functional the space of initial conditions that will equilibrate to an RJ spectrum with one (or two) condensate(s). Comparing the size of this set to the set of initial conditions that does not lead to condensation is outside the scope of this study.

\subsection{Lack of condensation in the infinite box limit}
\label{subsec:InfiniteBoxLimit}

In figures~\ref{fig:Condensation_convergence}(a)(ii)-(v), we see that the boundary layer of $\delta\mu$ values where condensates are forming gets pushed to progressively smaller values as we decrease $\epsilon$ (move down the panels within each triptych). It appears in figures~\ref{fig:Condensation_convergence}(a)(i) and (vi) as if the zonostrophy condensate survives in the $\epsilon\to0$ limit, however figures~\ref{fig:Condensation_convergence}(b)(i) and (vi) reveal that approaching the same points on the RJ boundary along lines of constant $\mu$ again sends the boundary layer to progressively smaller $\delta\lambda$ as we decrease $\epsilon$. Examining figure~\ref{fig:mulambda}(b), we see that the orange parts of the RJ boundary with $\mathbf{k}^+_-=(\epsilon,\epsilon)$ become increasingly parallel to lines of constant $\lambda$ as $\epsilon\to0$, i.e.\ the  $\delta\mu$ axes in figures~\ref{fig:Condensation_convergence}(a)(i) and (vi) remain within the boundary layer even for large $\delta\mu$. Likewise the $\delta\lambda$ axes in figures~\ref{fig:CHMphasediag_discr}(b)(ii)-(iv) also remain in the boundary layer due to the shallowness of the gradient of the RJ boundary in these cases.

We conclude that when we approach the RJ boundary in a perpendicular direction, transverse to the boundary layer (figures~\ref{fig:Condensation_convergence}(b)(i), (vi) and (a)(ii)-(v)), we see that the boundary layer becomes progressively narrower as $\epsilon\to0$ (although the boundary layer for the dominant invariant shrinks slower than that of the subdominant invariants). This is to say that condensation within the RJ spectrum does not remain in the infinite box limit, entirely in agreement with the Mermin-Wagner-Hohenberg theorem.

\section{CHM phase diagram}
\label{sec:CHMphasediag_discr}

\begin{figure}
	\centering
		\includegraphics[width=\textwidth]{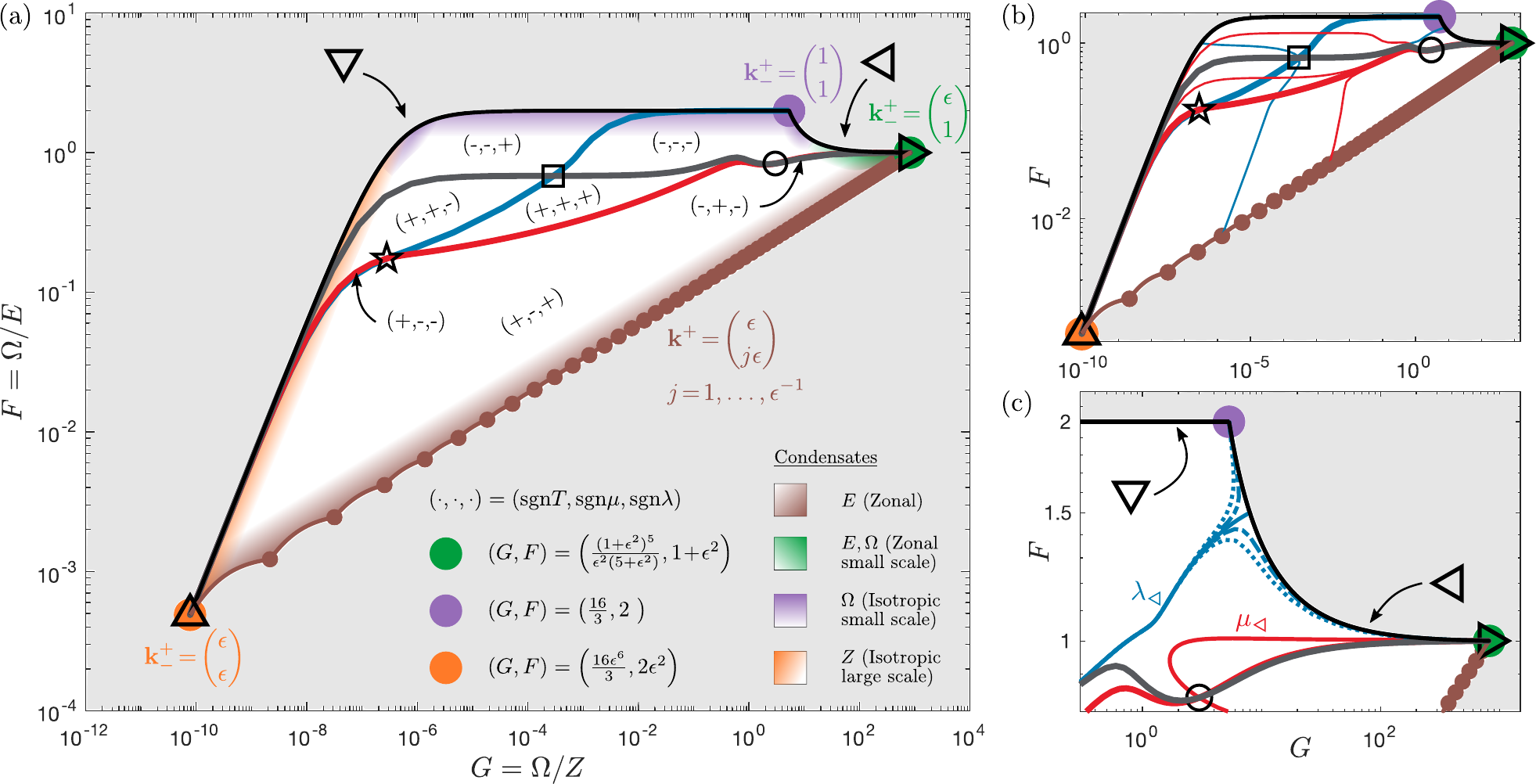}
		\caption{
			CHM phase diagram for $\epsilon \!=\!1/64$. 
			(a) Sectors of the phase diagram are marked according to $(\sgn T, \sgn\mu, \sgn\lambda)$, 
				partitioned by the lines $\lambda\!=\!0$ (thick blue), $\mu\!=\!0$ (thick red),  and $T,\mu,\lambda\!\to\!\infty$ (thick grey).   
				Shaded areas suggest the boundary layer where one or two dominant invariants condense first.
			(b) The phase diagram with the RJ sail traversed by lines of constant $\lambda$ (blue) and constant $\mu$ (red).
			(c) Upper right portion of the phase diagram, showing the lines
				$\lambda=\lambda_\medtriangleleft$ (solid blue), $\lambda_\medtriangleleft\pm 4\!\times\!10^{-5}$ (dashed blue),  
				$\lambda_\medtriangleleft\pm 12\!\times\!10^{-5}$ (dotted blue), and $\mu=\mu_\medtriangleleft$ (solid red).
		}
	\label{fig:CHMphasediag_discr}
\end{figure}

	By specifying the initial spectrum of the system we can prescribe the amount of $(\Omega,E,Z)$ present. These invariants retain their value as the spectrum evolves to equilibrium. If it is possible to accommodate the initial invariants in an RJ spectrum, then the three thermodynamic parameters $(T,\mu,\lambda)$ will have values that yield $(\Omega,E,Z)$ via Eqs.~\eqref{eq:invariants_RJ_discr}. We can eliminate the temperature $T$ by  considering the ratios 
\begin{equation}
\label{eq:GandF}
	G(\mu,\lambda)=\frac{\Omega}{Z}
		\qquad \mathrm{and}  \qquad
	F(\mu,\lambda)=\frac{\Omega}{E}.
\end{equation}
Viewing $G$ and $F$ as control parameters that we can manipulate through the initial spectrum, we can consider what subset of $G$--$F$ space corresponds to initial conditions that equilibrate to an RJ spectrum. We find this subset by varying $\lambda$ and $\mu$ over their allowed values for physical spectra, i.e.\  fixing $\lambda$ and sweeping $\mu$ through $(\mu^+\!(\lambda),\infty) \cup (-\infty,\mu_-\!(\lambda))$, then fixing $\mu$ and sweeping $\lambda$ through $(\lambda^+\!(\mu),\infty) \cup (-\infty , \lambda_-\!(\mu))$. This allows us to calculate $G$ and $F$ via equations~\eqref{eq:invariants_RJ_discr}, and  map out the sail-shaped region shown in figure~\ref{fig:CHMphasediag_discr}. 
	The ``RJ sail'' shows the range of control parameters that equilibrate to different kinds of RJ spectra, and thus can be viewed as the phase diagram of weakly nonlinear CHM turbulence. Figure~\ref{fig:CHMphasediag_discr} is plotted for $\epsilon=1/64$ but is qualitatively simular for general $\epsilon$.

\subsection{Interior of the RJ sail}
\label{subsec:RJsail_interior}

In figure~\ref{fig:CHMphasediag_discr}(a) we show that the  RJ ``sail'' is divided into different sectors by the thick blue line $\lambda=0$, the thick red line $\mu=0$, and the thick grey line. The latter is drawn by setting $|\lambda|=\mathrm{const}\gg1$ and sweeping through the allowed $\mu$; as we send $|\lambda|\to\infty$ the line collapses onto the grey line shown, which coincides with the line drawn by setting $|\mu|=\mathrm{const}\gg1$, sweeping $\lambda$, and sending $|\mu|\to\infty$. On this line in order to retain nonvanishing $(\Omega,E,Z)$ we need to also send $T\to\infty$, i.e.\ the grey line represents all of $(T,\mu,\lambda)\to\infty$.

	Thus, the sectors in figure~\ref{fig:CHMphasediag_discr}(a) each contain different combinations positive and negative $(T, \mu,\lambda)$ (with the $(+,-,-)$ and $(-,+,-)$ sectors lying in the indicated regions, too small to visualise at this scale).	
	Note that by varying $(G,F)$ continuously one can pass smoothly through the $\mu=0$, $\lambda=0$, and $T\to\infty$ lines continuously with no dramatic phase change, c.f.~\citep{fox1973inviscid, kraichnan1975statistical}. The latter line separates the $T>0$ sectors below from the $T<0$ sectors above, which comports with the interpretation of negative temperature states being ``hotter than any positive temperature''~\citep{purcell1951nuclear, kraichnan1975statistical, onorato2020negative}.

	The coloured shading in figure~\ref{fig:CHMphasediag_discr}(a) suggests schematically the areas near the edge of the RJ sail where we expect invariants to condense, following the arguments of section~\ref{sec:condensation}.  We indicate only the condensation of the invariant(s) associated with the fundamental mode $\mathbf{k}^+_-$ that dominates the behaviour near the edge of the sail. There will also be a narrower layer adhering to the edge of the sail (not shown in figure~\ref{fig:CHMphasediag_discr}(a)) where all three invariants condense.
	
	Important points of reference in the interior of the RJ sail are the point $\medwhitestar$ (representing $\lambda\!=\!\mu\!=\!0$); and the points $\medsquare$ (for $\mu\!=\!\pm\infty$), which is an accumulation point for all lines of constant $\lambda$ at their $\mu\to\pm\infty$ limit; and $\medcircle$ (for$\lambda\!=\!\pm\infty$)  which accumulates all lines of constant $\mu$ at their $\lambda\to\pm\infty$ limit. The accumulation of constant $\lambda$ and $\mu$ lines at $\medsquare$ and $\medcircle$ is shown in figure~\ref{fig:CHMphasediag_discr}(b) where we plot several lines of constant $\lambda$ and $\mu$ to chart out the coordinate lines in the interior of the RJ sail.
	
\subsection{Boundary of the RJ sail}
\label{subsec:RJsail_boundary}

The RJ boundary in the $G$--$F$  plane is defined by the singular limits of the RJ spectrum~\eqref{eq:nRJdelta} and~\eqref{eq:nRJtwodeltas}. In particular when the spectrum becomes a single Kronecker delta~\eqref{eq:nRJdelta}, the entire segment of the RJ boundary in $\lambda$--$\mu$ space corresponding to that $\mathbf{k}^+_-$  collapses to a point in $G$--$F$ space. These are the orange, green, purple, and brown points in figure~\ref{fig:CHMphasediag_discr}, labelled with their respective $\mathbf{k}^+_-$ in figure~\ref{fig:CHMphasediag_discr}(a). The $\epsilon$ dependences of the first three points are also given in the figure; we see that as $\epsilon\to0$ the RJ sail expands to cover the entire part of $G$--$F$ space with $F\leq 2$ (limited from above by the entire spectrum condensing into $\mathbf{k}_-=(1,1)$).

	The orange and green points also coincide respectively with the $\medtriangleup$ and $\medtriangleright$ points that mark the transition between different $\mathbf{k}^+$ regimes, as per~\eqref{eq:kPlus}. This reflects the fact that each $\mathbf{k}^+$ transition happens at a single mode, $(\epsilon,\epsilon)$ and $(\epsilon,1)$ respectively.

	By contrast, the transitions $\medtriangleleft$ and $\medtriangledown$ each involve two different $\mathbf{k}_-$  modes, see~\eqref{eq:kMinus}, so the spectrum at these points becomes the sum of two Kronecker deltas~\eqref{eq:nRJtwodeltas}. Their relative weighting $\alpha$ will depend on the direction of approach to the RJ boundary: along a line of constant $\mu$, a line of constant $\lambda$, or some other angle. This is demonstrated in figure~\ref{fig:CHMphasediag_discr}(c) for the $\medtriangleleft$ point. In the $\lambda$--$\mu$ plane the lines $\lambda = \lambda_\medtriangleleft$ and  $\mu=\mu_\medtriangleleft$ both approach the \emph{same} point $\medtriangleleft$ on the RJ boundary, but from different directions,  (consequently their respective approach to the $\medtriangleleft$ point favours $\mathbf{k}_-=(\epsilon,1)$ or $(1,1)$ in different amounts). However the images of these lines in the $G$--$F$ plane terminate on the RJ boundary at \emph{different} points.
Changing the angle of approach to $\medtriangleleft$ in $\lambda$--$\mu$ space changes the weighting $\alpha \in [0,1]$ in~\eqref{eq:nRJtwodeltas}, which sweeps out the black line labelled $\medtriangleleft$ in figure~\ref{fig:CHMphasediag_discr} that forms the upper right boundary of the RJ sail. 

We also plot the lines $\lambda=\lambda_\medtriangleleft \pm \Delta\lambda, 4\Delta\lambda$, with $\Delta\lambda = 4\!\times\!10^{-5}$, to show that lines either side of $\lambda_\medtriangleleft$ terminate at the purple ($+$ sign) and green ($-$ sign) points representing their respective $\mathbf{k}_-$ modes.  Viewed as a dynamical system, the $\lambda_\medtriangleleft$ trajectory is the separatrix between the basins of attraction of the purple and green stable fixed points, which attract constant $\lambda$  trajectories.

Similarly the entire upper left boundary (black in figure~\ref{fig:CHMphasediag_discr}) is formed by the spectrum linearly interpolating betwen the $\mathbf{k}_-$ modes at the $\medtriangledown$ point, and the lower right boundary (brown) is formed by the $(\epsilon^{-1}-1)$ interpolations between the successive zonal modes $\mathbf{k}^+=(\epsilon, j\epsilon)$, $j=1,2,\ldots ,\epsilon^{-1}$.

As mentioned in section~\ref{sec:condensation}, as we approach the boundary of the RJ sail, the denominator of $\nRJ$ is shrinking to zero, so in order to keep $(\Omega,E,Z)$ finite, $T$ must also vanaish. In figure~\ref{fig:CHMphasediag_discr} we must have $T\to0^+$ as we approach the brown boundary, and $T\to0^-$ as we approach either black boundary.

\subsection{Outside the RJ sail}
\label{subsec:RJsail_exterior}

\begin{figure}
	\centering
		\includegraphics[width=0.8\textwidth]{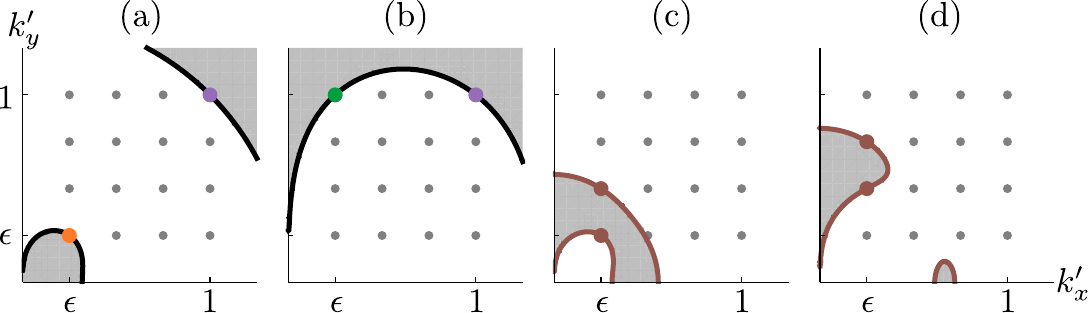}
		\caption{
			$\sgn P(\alpha,\mathbf{k}_1,  \mathbf{k}_2, \mathbf{k'})$ over $\mathbf{k'}$ space for $L_\epsilon=1/4$, $\alpha=1/2$. 
			We choose $\mathbf{k}_1$ and $\mathbf{k}_2$ so that $P$ represents 
			perturbations of the spectrum away from boundaries of the RJ sail in figure~\ref{fig:CHMphasediag_discr}: 
			(a) $\medtriangledown$ black boundary, (b) $\medtriangleleft$ black boundary, 
			(c) first segment of the lower brown boundary, (d) second segment of the lower brown boundary.
			The thick black or brown lines are contours of $P=0$, white areas denote $P>0$, and grey areas $P<0$.
		}
	\label{fig:PertnEdge}
\end{figure}

In section~\ref{subsec:InfiniteBoxLimit} we noted that condensates within the RJ spectrum only exist at finite size and do not remain in the $\epsilon\to0$ limit. Equivalently, in this limit any spectrum initialised over the lattice $L_\epsilon$ equilibrates to an RJ spectrum with some physically meaningful $(T,\mu,\lambda)$. A related question is, for finite $\epsilon$, whether singular spectra could exist that contain a macroscopic amount of invariants, but which lie outside the RJ spectrum. If such a spectrum exists, this possibility would be similar to Bose-Einstein condensation in the 3D GPE in a spatially infinite system, where below the condensation temperature excess particles are absorbed into the singular spectrum $\delta(\mathbf{k})$.
Equivalently we can ask whether there are initial spectra that contain combinations of $(\Omega,E,Z)$  that cannot be accommodated by any $(T,\mu,\lambda)$ in an RJ spectrum. This amounts to asking whether there is any physical spectrum that maps to a point outside the RJ sail. 

To address this we note that in the discrete description, any arbitrary but physically meaningful spectrum is a non-negative function over $L_\epsilon$, i.e.\ it is a linear combination of Kronecker deltas on each lattice point: $n_\mathbf{k}= \sum_{\mathbf{k'}\in L_\epsilon}a_\mathbf{k'}\delta(\mathbf{k}-\mathbf{k'})$. In particular, any spectrum that maps to a point outside the RJ sail in the $G$--$F$ plane must also have this structure. We therefore proceed by choosing a spectrum corresponding the boundary of the RJ sail and examining all possible small perturbations
\begin{equation}
	\label{eq:perturbed_spectrum}
	n^{\scriptscriptstyle \mathrm{RJ (bdry)}}_\mathbf{k} 
	\to 	n^{\scriptscriptstyle \mathrm{RJ (bdry)}}_\mathbf{k}   +  \eta \delta(\mathbf{k}-\mathbf{k'}),
\end{equation} 
where $\eta\ll 1$, $\mathbf{k'}\in L_\epsilon$, and $n^{\scriptscriptstyle \mathrm{RJ (bdry)}}_\mathbf{k}$ is the limiting spectrum~\eqref{eq:nRJtwodeltas} on the RJ boundary. (We do not consider the limit~\eqref{eq:nRJdelta} as the observations of section~\ref{subsec:RJsail_boundary} show that	such spectra map to a finite number of isolated points in the $G$--$F$ plane.) 
If all such small perturbations to the spectrum were to map to points in the  $G$--$F$ plane that are inside the RJ sail, we conjecture that any finite perturbation would map to a points even further inside the sail. We further reason that any arbitrary combination of finite perturbations from the boundary would also land inside the RJ sail, i.e.\ \emph{any} physical spectrum would be represented inside the RJ sail in $G$--$F$ space.

First we recall that the black and brown boundary lines of the RJ sail represent systems with spectra given by~\eqref{eq:nRJtwodeltas}, i.e.\ a sum of two Kronecker deltas. For any of these boundaries, a point parametrised by $\alpha$ has e.g.\ zonostrophy
\begin{equation*}
	Z^\delta = \alpha \varphi_1 + (1-\alpha) \varphi_2,
\end{equation*}
and similarly enstrophy 
$\Omega^\delta$ and energy $E^\delta$ with  $\varphi_\mathbf{k}$ replaced by $ k_x$ and $\omega_\mathbf{k}$ respectively.\footnote{
		In this section superscripts label the character of spectra, invariants, etc.\ and 
		no longer refer to quantities on the $T>0$ branch of the RJ spectrum. 
		Nor do they imply any summation over tensor indices.}

This point on the RJ boundary in $G$--$F$ space has coordinates $G^\delta=\Omega^\delta/Z^\delta$ and $F=\Omega^\delta/E^\delta$. The tangent vector in the $G$--$F$ plane is then $(\partial_\alpha G^\delta, \partial_\alpha F^\delta)$, and the inward pointing normal vector 
\begin{equation*}
	\mathbf{N}^\delta = \pm \left( \partial_\alpha F^\delta , - \partial_\alpha G^\delta \right)
\end{equation*}
is selected by choosing the appropriate sign.

We then perturb the spectrum by adding introducing Kronecker delta at $\mathbf{k'}$ as in~\eqref{eq:perturbed_spectrum} and calculate the invariants 
\begin{equation*}
 	Z^\eta = \alpha \varphi_1 + (1-\alpha) \varphi_2 + \eta\varphi_\mathrm{k'}
\end{equation*}
etc, then take ratios to find $G^\eta$ and $F^\eta$. The vector 
\begin{equation*}
	\mathbf{X}^\eta = \left( \partial_\eta G^\eta,  \partial_\eta F^\eta\right)_{\eta=0}
\end{equation*}
then indicates the direction that the perturbation~\eqref{eq:perturbed_spectrum} takes the point lying on the RJ boundary in the $G$--$F$ plane.
Finally, the sign of the quantity
\begin{equation*}
	P(\alpha, \mathbf{k}_1,\mathbf{k}_2, \mathbf{k'}) = \mathbf{N}^\delta  \cdot \mathbf{X}^\eta 
\end{equation*}
determines whether the perturbation~\eqref{eq:perturbed_spectrum} takes the $(G,F)$ point into the RJ region ($P>0$) or outside it ($P<0$).

In figure~\ref{fig:PertnEdge} we plot $\sgn P(\alpha,\mathbf{k}_1,\mathbf{k}_2, \mathbf{k'})$ over $\mathbf{k'}$ space, for $\mathbf{k}_1$, $\mathbf{k}_2$ corresponding to: (a) the upper left black boundary $\medtriangledown$ of the RJ sail; (b) the upper right black boundary $\medtriangleleft$; (c) the first segment of the lower right brown boundary with $\mathbf{k}_1=(\epsilon,\epsilon)$, $\mathbf{k}_2=(\epsilon,2\epsilon)$; and (d) another segment of the lower right brown boundary with $\mathbf{k}_1=(\epsilon,j\epsilon)$, $\mathbf{k}_2=(\epsilon,(j+1)\epsilon)$; here $j\!=\!2$ but the diagram is similar for any $1\!<\!j\!<\!(\epsilon^{-1}\!-\!1)$. The black and brown contours mark $P=0$, the white regions denote $P>0$ and the grey regions $P<0$. 
In figures~\ref{fig:PertnEdge}(a)-(d) we have chosen $\epsilon=1/4$ for clarity, but the figures are qualitatively similar as we decrease $\epsilon$. We plot the figures with $\alpha=1/2$; as we change $\alpha$ the contours of $P=0$ do not change appreciably, however the contours inside $P>0$ and $P<0$ (not shown in figure~\ref{fig:PertnEdge} for clarity) change scale rapidly with $\alpha$. 

In all of figures~\ref{fig:PertnEdge}(a)-(d), we observe that $P\geq0$ for each $\mathbf{k'}\in L_\epsilon$. We note that in (c) the $P<0$ region isolates the lattice point $\mathbf{k'}=(\epsilon,\epsilon)$ from the rest of $L_\epsilon$, however in all cases we have examined, no lattice point lies within the $P<0$ region. We thus conclude that no infinitesimal perturbation of the spectrum away from $n^{\scriptscriptstyle \mathrm{RJ (bdry)}}_\mathbf{k}$  can reach a point outside the RJ sail pictured in figure~\ref{fig:CHMphasediag_discr}. By the argument above we conjecture that no finite perturbation from~\eqref{eq:nRJtwodeltas}, i.e.\ no physically allowable spectrum corresponds to a point outside the sail: every physical spectrum will thermalise to an RJ distribution.

\section{Discussion and conclusion}
\label{sec:conclusion}

We have carried out a comprehensive study of the equilibrium RJ spectra of the CHM equation in its WT limit, in a double Fourier-truncated system. We have identified the range of thermodynamic parameters $\lambda$ and $\mu$ that correspond to physical equilibria, and the fundamental modes $\mathbf{k}^+_-$ at which the RJ spectrum becomes singular and condense significant fractions of the dynamical invariants. 
	The presence of three invariants widens the variety of fundamental modes that can accumulate condensates, as compared to the 2D Euler and GPE cases with two invariants. In addition to condensates at the gravest (largest-scale) and smallest-scale mode, as in the two-invariant case, in the three-invariant CHM system we have condensates at every zonal mode. 

We have found that the invariant (or invariants) that condense at each fundamental $\mathbf{k}^+_-$ is the invariant whose density dominates in the sector where that $\mathbf{k}^+_-$ is located. Thus,  $\Omega$ condenses into the smallest-scale modes, either isotropic or zonal; $E$ condenses into zonal modes; and $Z$ into the gravest isotropic mode. We have thus an explanation for the turbulent construction of ordered structures such as zonal jets, and large and small vortices in quasi-2D flows with a $\beta$ effect, in terms of the equilibria sought by the turbulence. This explanation is the equilibrium equivalent of the the Fj{\o}rtoft argument regarding how the condensates will form dynamically via the cascade of invariants~\citep{nazarenko2009triple}. Indeed the two scenarios work hand in hand, with dynamical cascades bringing the system from its initial condition to an RJ equilibrium. 

Furthermore, we can speculate on the subsequent evolution if there is weak dissipation in the system, for example viscous processes at the smallest scale. The Fj{\o}rtoft argument can be expressed in terms of the centroids of invariants: scales that characterise the location of the invariants in $\mathbf{k}$-space. If the spectrum changes, it does so in such a way that the centroid of each invariant moves deeper into its respective sector~\citep{nazarenko2009triple, nazarenko2011waveturbbook}. The leak of $\Omega$ from the system at high $k$ due to viscous dissipation would simultaneously push the $E$ and $Z$ centroids deeper into their sectors. Dissipation at high $k$ would thus strengthen zonal or large-scale isotropic condensates. A similar process has been suggested for the dynamic formation of Bose-Einstein condensates via evaporative cooling in the GPE~\citep{nazarenko2011waveturbbook}.

We note that the condensates we have described in this paper do not survive in the infinite box limit. However, the large-scale character of Rossby (drift) waves and zonal flows is such that their size is comporable to the planetary atmosphere or ocean (fusion device) in which they propagate. Thus the examination we have carried out here of finite-sized systems remains practically relevant.

This analytical study leaves open some questions regarding the dynamical relevance of condensates, in terms of relating the width of the ``boundary layer'' near the RJ edge to the volume of initial conditions that can equilibrate to a condensed state. This is to say, whether it is necessary to start with an initial condition that is already similar to an RJ spectrum with a condensate, in order to realise such a state in equilibrium, or whether a wider range of initial conditions will evolve into a condensed state. Careful characterisation of this question would require relating physically relevant norms in the functional space of initial conditions to the area of the boundary layer in the $\lambda$--$\mu$ plane.

We also have not related states of negative thermodynamic potential to entropy derivatives, as has been recently done for GPE-like systems~\citep{onorato2020negative, baudin2021energy}. And obviously, our predictions remain to be tested by numerics. We leave these challenges open to future work.

Finally, we note that the three-invariant case considered here arises in the small-scale limit of the CHM $\rho k\gg1$. In the large-scale limit $\rho k \ll 1$, which is more relevant to fusion plasmas, there is a fourth invariant~\citep{saito2013angular}, termed the semi-action~\citep{ConnNazQuinn2015rossby}. The Fj{\o}rtoft argument was extended to this four-invariant case by Harper and Nazarenko~\citep{harper2016large}. Classifying the condensation behaviour in this case, using the methodology we have developed here, is an obvious next step in fully characterising Rossby and drift wave turbulence.

\ack
JS thanks Colm Connaughton for many useful discussions and invaluable support in the preparation of this manuscript.
SN is supported by the Chaire D'Excellence IDEX (Initiative of Excellence) awarded by Universit{\'e} de la C{\^o}te d'Azur, France, the EU Horizon 2020 research and innovation programme under the grant agreements No 823937 in the framework of Marie Skodowska-Curie HALT project and No 820392 in the FET Flagships PhoQuS project and the Simons Foundation Collaboration grant `Wave Turbulence' (Award ID 651471).
JS is supported by the UK EPSRC through Grant No EP/I01358X/1.

\appendix

\section{Application to the Gross-Pitaevskii equation}
\label{app:GPE}

Here we summarise the results of the GPE in spatially infinite systems, and state how our results on the finite-sized CHM apply to the GPE. 

\subsection{Review of the GPE in an infinite-sized domain}

The Gross-Pitaevskii equation~\citep{gross_structure_1961, pitaevskii_vortex_1961},
\begin{equation}
\label{eq:GPE}
i \frac {\partial \psi}{\partial t} + \nabla^2 \psi + s\left| \psi \right|^2 \psi =0,
\end{equation} 
is a classical model used to study condensation in Bose gases~\citep{pitaevskiistringari2003book}, and also in light propagating in a dispersive medium with a local Kerr nonlinearity~\citep{dyachenko1992wave} (in the optical case the dimensionality $d<3$). In the Bose gas case $\psi(\mathbf{x},t)$ is the boson wavefunction, and in the optical case it represents the modulating envelope of the input beam. We condsider the defocusing GPE with $s=-1$, in which case condensates are stable.

Proceeding as in section~\ref{subsec:CHM_WT_KE}, we work in Fourier space and define the waveaction spectrum $n_\mathbf{k} = \left(\frac{L}{2\pi}\right)^d\langle|\psi_\mathbf{k}|^2\rangle$. We make the standard WT assumptions, which allow us to derive a four-wave kinetic equation, describing the evolution of  $n_\mathbf{k}$ for small-amplitude waves with dispersion relation $\omega_\mathbf{k}=k^2$ fluctuating over a zero background~\citep{dyachenko1992wave}: 
\begin{equation*}
		\partial_t n_\mathbf{k}  = 4
			\pi   \!  \int  \! 
			 \ n_1 n_2 n_3 n_\mathbf{k}   
			\left[\frac{1}{n_\mathbf{k}} + \frac{1}{n_3} - \frac{1}{n_1} - \frac{1}{n_2} \right]
			\delta^{12}_{3\mathbf{k}} \delta(\omega^{12}_{3\mathbf{k}}) 
		 	\, \mathrm{d}\mathbf{k}_1  \mathrm{d}\mathbf{k}_2   \mathrm{d}\mathbf{k}_3,
\end{equation*}
where $\delta^{12}_{3\mathbf{k}}=\delta(\mathbf{k}_1+\mathbf{k}_2-\mathbf{k}_3-\mathbf{k})$ and
$\delta(\omega^{12}_{3\mathbf{k}}) = \delta(\omega_1+\omega_2 - \omega_3 - \omega_\mathbf{k})$.

This kinetic equation preserves $\mathcal{N}$, the total waveaction, or number of particles in the boson case, and the linear energy $\mathcal{E}$,
\begin{equation}
\label{eq:GPE_NandE}
\mathcal{N}=\int \! n_\mathbf{k} \, \mathrm{d}\mathbf{k}
	\qquad \qquad 
	\mathrm{and} 
	\qquad \qquad
\mathcal{E} =\int \! \omega_\mathbf{k} n_\mathbf{k} \, \mathrm{d}\mathbf{k},
\end{equation} 
and has a class of stationary solution that are the equilibrium RJ spectra
\begin{equation}
\label{eq:GPE_RJ}
\nRJ=\frac{T}{\omega_\mathbf{k}+\mu}
\end{equation}
where $T$ is the temperature and  $\mu$ is the chemical potential. 

On the RJ spectrum~\eqref{eq:GPE_RJ} the invariant $\sigma=(\mathcal{E}+\mu\mathcal{N})/T$ is partitioned equally across all Fourier modes. Again this leads to the ultraviolet catastrophe if Fourier space is unbounded. In~\citep{connaughton2005condensation} this is regularised by introducing a radial cutoff $k_c$ to the $\mathbf{k}$-space integrals in~\eqref{eq:GPE_NandE}. This hard cutoff is more physically motivated in  \citep{nazarenko2011waveturbbook} for the case of particles obeying Bose-Einstein statistics, with distribution $n_\mathbf{k} = (\exp[(k^2+\mu)/T ]- 1)^{-1}$. The RJ spectrum \eqref{eq:GPE_RJ} is the low-$k$ limit of the Bose-Einstein distribution, whereas at high $k$ the distribution becomes a Gaussian cutoff $n_\mathbf{k} \sim \exp(-k^2/T)$. Whether hard or exponential, the ultraviolet cutoff arrests the runaway excitation of high wavenumbers during thermalisation, allowing \eqref{eq:GPE_RJ} to become a nontrivial stationary spectrum. 

For spatially infinite domains there is no low-$k$ cutoff, and~\eqref{eq:GPE_NandE} remain as continuous integrals.

\subsection{RJ spectra and Bose-Einstein condensation in 3D.}

\begin{figure}
	\centering 
	\includegraphics[width=0.999\textwidth]{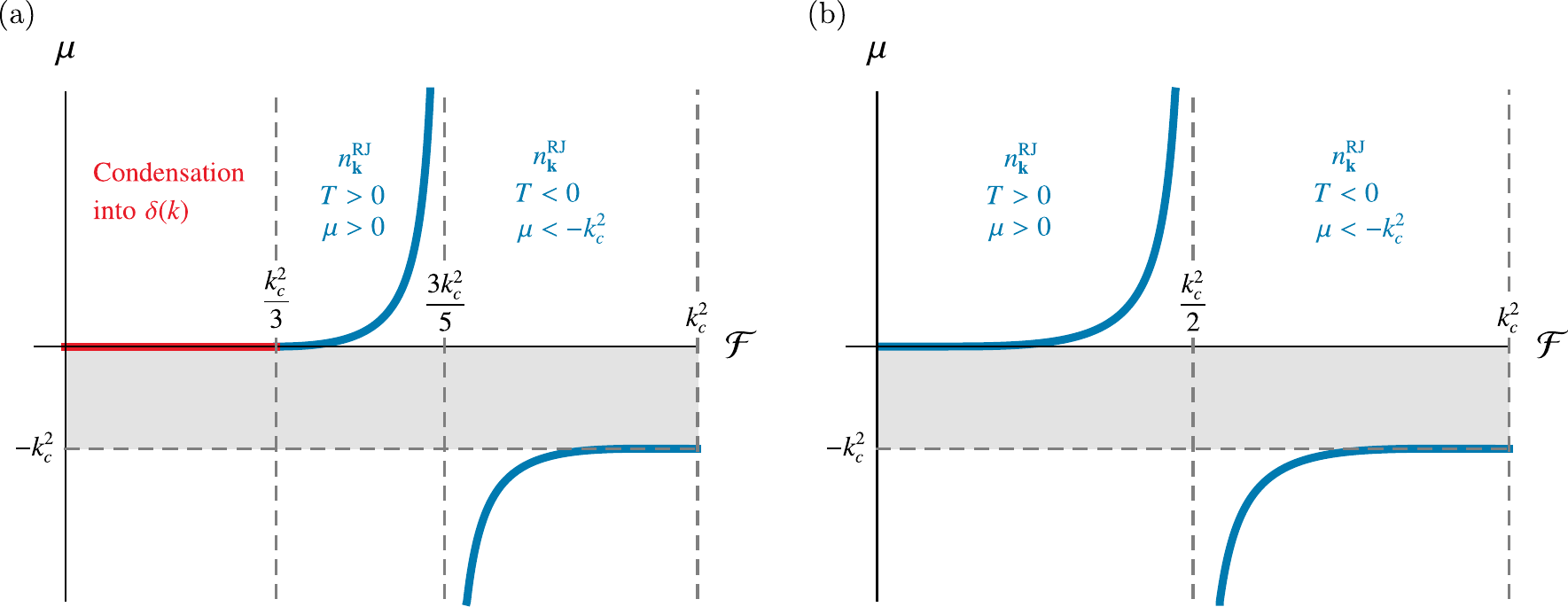}
	\caption{
			Chemical potential $\mu$ as a function of the energy per particle $\mathcal{F}=\mathcal{E}/\mathcal{N}$ 
			for an infinite-sized GPE system with high-wavenumber cutoff at $|\mathbf{k}|=k_c$, in (a) 3D, and (b) 2D. 
			For both dimensionalities the RJ spectrum has a positive and a negative $(T,\mu)$ branch. 
			Condensation into a singular spectrum happens in 3D but not in 2D.
			}
	\label{fig:GPEinfSizePolar}
\end{figure}
	
We consider a 3D system initialised with some amount of waveaction and energy, and evolving to an equilibrium state. If this equilibrium can be described by~\eqref{eq:GPE_RJ} then $\mu$ and $T$ will be defined in terms of the initial $\mathcal{N}$ and $\mathcal{E}$, which in the spatially infinite system with a $\mathbf{k}$-space cutoff at $k_c$ are
\begin{equation}
	\label{eq:GPE_NandE_RJ_3d}
	\mathcal{N} =  4\pi T \int_0^{k_c} \!     \frac{k^2}{k^2+\mu}       \, \mathrm{d}k
	\qquad \qquad
	\mathcal{E} =  4\pi T \int_0^{k_c} \!    \frac{k^4}{k^2+\mu}    \, \mathrm{d}k.
\end{equation}
These integrals can be evaluated explicitly, see~\citep{connaughton2005condensation} for their closed-form expression. 
For $\mu>0$ at fixed $T>0$, both $\mathcal{N}$ and $\mathcal{E}$ monotonically decrease with $\mu$. At $\mu=0$ both integrands vary as $k^\alpha$ with $\alpha \geq 0$, so both $\mathcal{N}$ and $\mathcal{E}$ converge at the lower limit in 3D. Thus for every $T>0$ the values of $(\mathcal{N},\mathcal{E})$ that can be accommodated in the RJ spectrum are bounded from above by their value at $\mu=0$. 

We define the energy per particle
\begin{equation*}
	\mathcal{F}(\mu) = \frac{\mathcal{E}}{\mathcal{N}},
	\end{equation*} 
which is independent of $T$, and can be considered a control parameter in numerical simulations by choosing the initial spectrum appropriately. Following~\citep{gallet2015WT_KGMH} we plot the $\mu$ that characterises the equilibirum spectrum as a function of the initial $\mathcal{F}$, see figure~\ref{fig:GPEinfSizePolar}(a).

As $\mu\to 0^+$, $\mathcal{N}$ and $\mathcal{E}$ increase to their local maxima while their ratio decreases to a critical value $\mathcal{F}(0)$, which in the case of a radial cutoff equals $k_c^2/3$. Systems initialised with $\mathcal{F}< \mathcal{F}(0)$ (red region in figure~\ref{fig:GPEinfSizePolar}(a)) cannot be accommodated by the RJ spectrum. Einstein's argument~\citep{einstein1925aQuantentheorie} is that once $\nRJ$ reaches its capacity at $\mu=0$ any extra particles are absorbed into the zero-energy state, i.e.\ into a singular distribution at $\mathbf{k}=0$. This is the true Bose-Einstein condensate, and can exist at nonzero temperature.    

At the critical value $\mathcal{F}\!=\!0$ for condensation, $\mu\!=\!0$ and the RJ spectrum represents equipartition of $\mathcal{E}$. Above the condensation threshold, systems initialised with $k_c^2/3 < \mathcal{F} < 3k_c^2/5$ equilibrate to an RJ spectrum with no condensate, and positive $(\mu,T)$. 

As $\mathcal{F} \to (3k_c^2/5)^\mp$ we have $\mu\to\pm\infty$. For finite $(\mathcal{N},\mathcal{E})$ this means that $T \to \pm\infty$, and the RJ spectrum is one of equipartition of $\mathcal{N}$.

	For $\mathcal{F}>3k_c^2/5$  physical RJ solutions exist with $T < 0$ and $-\infty<\mu<-k_c^2$, in the sense that $\nRJ>0$ for all $\mathbf{k}: 0\leq k \leq k_c$. These negative temperature RJ states are characterised by $n_\mathbf{k}$ rising with $k$, becoming increasingly peaked as $\mu$ increases from $-\infty$. As $\mu\to (-k_c^2)^-$ and  $\mathcal{F}\to k_c^2$ we must have $T\to 0^-$ so as to keep $(\mathcal{N},\mathcal{E})$ finite. Eventually the RJ spectrum becomes singular $\nRJ \to \mathcal{N} \delta(k-k_c)/4\pi k^2$. As this limit is approached, $\mathcal{E}$ and $\mathcal{N}$ get increasingly concentrated into spherical shells of decreasing thickness $|k_c-k|\to0$ (with the $\mathcal{E}$ shell being thicker than the $\mathcal{N}$ shell). This is condensation in the sense that we defined in section~\ref{subsubsec:CHM_RJ}, extended to continuous $\mathbf{k}$-space, i.e.\ macroscopic occupation of the invariants in a set of Fourier modes whose measure vanishes in the limit.
		
	For  $-k_c^2 <\mu<0$,  in between the positive and negative $(\mu,T)$ branches of $\nRJ$, there exists a gap in which no physical RJ solution is possible. This gap is shaded grey in figure~\ref{fig:GPEinfSizePolar}.

To summarise: in 3D, for all inital $(N,E)$ that are physically realisable, the GPE dynamics drive the system to equilibria that consist of either a condensate with some RJ component, an RJ spectrum with $(\mu,T)$ positive, or an RJ distribution with $(\mu,T)$ negative.

\subsection{RJ spectra in 2D with no true Bose-Einstein condensation}

For an infinite system in 2D with an isotropic cutoff at high wavenumber, the waveaction and energy on an RJ spectrum are
\begin{equation}
\label{eq:GPE_NandE_RJ_2d}
\mathcal{N} =  2\pi T \int_0^{k_c} \!     \frac{k}{k^2+\mu}       \, \mathrm{d}k
\qquad \qquad
\mathcal{E}=  2\pi T \int_0^{k_c} \!    \frac{k^3}{k^2+\mu}    \, \mathrm{d}k.
\end{equation}

Again for $T>0$ both $\mathcal{N}$ and $\mathcal{E}$ are maximised as $\mu=0$, when the integrands become $k^{-1}$ and $k$ respectively. The integral for $\mathcal{E}$ converges, indicating a finite capacity for the RJ spectrum to absorb energy at positive temperature. However the $\mathcal{N}$ integral diverges logarithmically at the lower limit, indicating that the RJ spectrum can accommodate an arbitrarily large number of particles. In order to fit this large $\mathcal{N}$ into the RJ spectrum, while keeping $\mathcal{E}$ constant we must have $T\to0^+$ as $\mu\to0^+$: in an infinite 2D system there is no true Bose-Einstein condensation of particles at finite temperature \citep{MerminWagner, Hohenberg}. 

This is reflected in the $\mu$ vs.\ $\mathcal{F}$ plot in figure\ \ref{fig:GPEinfSizePolar}(b). We have $\mu\to0^+$ as $\mathcal{F}\to0^+$, with no condensation (and the spectrum at $\mu=0$ representing equipartition of $\mathcal{E}$). In the range $0<\mathcal{F}<k_c^2/2$, the RJ spectrum has $(\mu,T)>0$ . In $k_c^2/2<\mathcal{F}<k_c^2$ the RJ spectrum has  $T<0$ and $-\infty<\mu<-k_c^2$; for such a spectrum $\nRJ$ increases with $k$.

As $\mathcal{F} \to (k_c^2/2)^\mp$ both $(\mu,T)\to\pm\infty$, and $\mathcal{N}$ is partitioned equally amongst wave modes. As $\mathcal{F}\to(k_c^2)^-$ and $\mu\to(-k_c^2)^-$ the spectrum reaches a singular limit, $\nRJ\to\mathcal{N}\delta(k-k_c)/2\pi k$ with $T\to0^-$ to keep the invariants finite. The density of $\mathcal{E}$ dominates that of $\mathcal{N}$ for such a spectrum. Again this is condensation in the sense of section~\ref{subsubsec:CHM_RJ}, extended to continuous integrals.

\subsection{The 2D GPE in a finite domain}

\begin{figure}
	\centering 
	\includegraphics[width=0.999\textwidth]{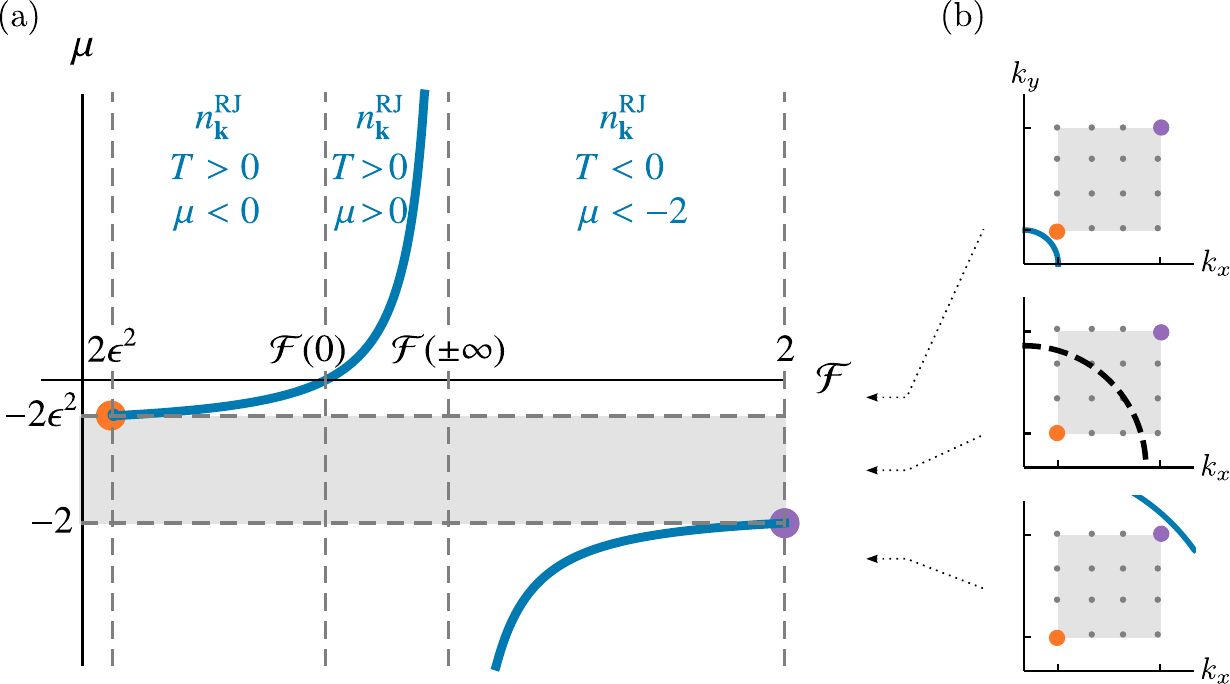}
	\caption{
			(a)	 Chemical potential $\mu$ vs.\ energy per particle $\mathcal{F}=\mathcal{E}/\mathcal{N}$ for a 2D GPE system of finite size,
				 with high and low wavenumber cutoffs in $k_x$ and $k_y$.
				 The introduction of a finite box size induces a portion of $\mu<0$ on the $T>0$ branch of the RJ spectrum. 
			(b) The curve of $D=0$ moving through the $k_x$--$k_y$ plane 
				as $\mu$ swings from the $\mu<0, T>0$ branch of $\nRJ$ (top panel), 
				into the RJ-forbidden region (middle), 
				and onto the $T<0,\mu<-2$ branch of $\nRJ$ (bottom).
			}
	\label{fig:GPEfiniteSize}
\end{figure}

We now turn to the GPE in a finite domain in 2D. We consider a system with the same setup as we used for the CHM equation in section~\ref{subsubsec:CHM_RJ}, with a high-wavenumber cutoff at $k_\mathrm{max}$ and low-wavenumber cutoff at $k_\mathrm{min}$ in both the $k_x$ and $k_y$ directions, the variables rescaled to $k_\mathrm{max}$ as in~\eqref{eq:CHM_rescale_vbles}, and the integrals for in~\eqref{eq:GPE_NandE} replaced by sums over the $\mathbf{k}$-space lattice $L_\epsilon$.
	We calculate $\mathcal{N}$ and $\mathcal{E}$ numerically to find $\mathcal{F}(\mu)=\mathcal{E}/\mathcal{N}$, and plot $\mu$ vs.\ $\mathcal{F}$ in figure~\ref{fig:GPEfiniteSize}(a). 

Once again we observe that the RJ spectrum has a positive $T$ branch and a negative $(\mu,T)$ branch, which are characterised by the spectrum falling or rising with $k$ respectively. Between the two branches $\mu\to\pm\infty$ as $\mathcal{F}$ approaches the limiting value $\mathcal{F}(\pm\infty)$ from below or above. Summing over the lattice $L_\epsilon$ we find $\mathcal{F}(\pm\infty)=(2+3\epsilon+\epsilon^2)/3$. As in the infinite-size case, this limit corresponds to equipartition of $\mathcal{N}$ with $T\to\pm\infty$. 

On the positive $T$ branch of $\nRJ$, we see that the introduction of the box size pushes the curve downwards in figure~\ref{fig:GPEfiniteSize}(a), so that there is now a $\mu<0, T>0$ part of the branch and a $(\mu,T)>0$ part. 
The $\mu<0$ part terminates as $\mathcal{F}\to2\epsilon^2$, with $\mu\to\mu^+=-2\epsilon^2$ and $T\to0$, with all limits approached from above. In this limit the spectrum becomes a Kronecker delta at the gravest mode $\mathbf{k}^+=(\epsilon,\epsilon)$, signified by the orange dot in figure~\ref{fig:GPEfiniteSize}. This is the condensate, as defined in section~\ref{subsec:GPE_condensation}. Here the waveaction density dominates over the energy density, so as this limit is approached $\mathcal{N}$ condenses first, with condensation of $\mathcal{E}$ following. This is exactly like the condensation of $Z$ in the CHM equation, except that in the GPE the absence of the third invariant leaves room for $\mathcal{N}$ to dominate in the entire sector $k\ll1$. 

The negative $T$ branch terminates at $\mathcal{F} \!=\! 2$  when the whole spectrum is concentrated at the  smallest-scale mode $\mathbf{k}_-=(1,1)$, signified by the purple dot. In this sector the energy density dominates over the waveaction density, so this limit represents condensation of primarily $\mathcal{E}$, and then finally $\mathcal{N}$ in the limit, as $\mu\to\mu^-=-2$ and $T\to 0$ from below (c.f.\ the condensation of $\Omega$ in the CHM equation).

In figure~\ref{fig:GPEfiniteSize}(b) we show the $k_x$--$k_y$ plane and how the curve $D=0$ changes (where $D(\mathbf{k},\mu)$ is the denominator of~\eqref{eq:GPE_RJ}) as $\mu$ is swung from the $T>0$ branch with $\mu^+<\mu<0$ (top panel), to inside the RJ-forbidden range  $\mu^+>\mu>\mu_-$ (middle panel), to the $T<0$, $\mu<\mu^-$ branch (bottom panel). We see that in the RJ-forbidden range of $\mu$ (grey region in figures~\ref{fig:GPEinfSizePolar} and~\ref{fig:GPEfiniteSize}(b)) , the $D=0$ line enters $S_\epsilon$, and ``captures'' modes of $L_\epsilon$, reversing the sign of the spectrum on these modes. Only when $D=0$ has captured all of $L_\epsilon$ does the spectrum become sign-definite again, on the $(\mu,T)<0$ branch.

\newcommand{\newblock}{}  
\bibliographystyle{unsrtnat}  

\bibliography{Condensn_in_QG_turb}

\end{document}